\documentclass[journal]{IEEEtran}
\usepackage{latexsym}

\IEEEoverridecommandlockouts
\usepackage{lipsum}
\usepackage{cite}
\usepackage[T1]{fontenc}
\usepackage[cmintegrals]{newtxmath}
\usepackage{array}
\usepackage[caption=false,font=footnotesize]{subfig}
\usepackage{url}
\usepackage{balance}
\usepackage{graphicx}
\usepackage{algorithm}
\usepackage{algpseudocode}
\usepackage{pifont}
\usepackage{amsmath}
\usepackage{float}
\usepackage{multicol}
\usepackage{mwe}
\usepackage{subfig}
\usepackage{epstopdf}
\usepackage[utf8]{inputenc}
\DeclareGraphicsExtensions{.eps}
\usepackage{array}
\usepackage{amsmath}
\usepackage{mathtools}
\usepackage[subnum]{cases}
\usepackage{color,soul}
\DeclareMathOperator*{\argmin}{argmin}

\DeclareMathOperator{\Tr}{Tr}

\makeatother
\begin{document}

\title{Wireless Networks with Cache-Enabled and Backhaul-Limited Aerial Base Stations}

\author{\IEEEauthorblockN{Elham Kalantari\IEEEauthorrefmark{1},
		Halim Yanikomeroglu\IEEEauthorrefmark{2}, and 
		Abbas Yongacoglu\IEEEauthorrefmark{1}\thanks{This work is supported in part by the Natural Sciences and Engineering Research Council (NSERC) of Canada and in part by Huawei Canada Co., Ltd.}}\\
	\IEEEauthorblockA{\IEEEauthorrefmark{1}School of Electrical Engineering and Computer Science\\
		University of Ottawa, Canada, email: \{ekala011, yongac\}@uottawa.ca}\\
	\IEEEauthorblockA{\IEEEauthorrefmark{2}Department of Systems and Computer Engineering\\
		Carleton University, Canada, email: halim@sce.carleton.ca}}


\maketitle

\begin{abstract}
	Use of aerial base stations (ABSs) is a promising approach to enhance the agility and flexibility of future wireless networks. ABSs can improve the coverage and/or capacity of a network by moving supply towards demand. Deploying ABSs in a network presents several challenges such as finding an efficient 3D-placement of ABSs that takes network objectives into account. Another challenge is the limited wireless backhaul capacity of ABSs and consequently, potentially higher latency incurred. Content caching is proposed to alleviate the backhaul congestion and decrease the latency. We consider a limited backhaul capacity for ABSs due to varying position-dependent path loss values and define two groups of users (delay-tolerant and delay-sensitive) with different data rate requirements. We study the problem of jointly determining backhaul-aware 3D placement for ABSs, user-BS associations and corresponding bandwidth allocations while minimizing total downlink transmit power. Proposed iterative algorithm applies a decomposition method. First, the 3D locations of ABSs are found using semi-definite relaxation and coordinate descent methods, and then user-BS associations and bandwidth allocations are optimized. The simulation results demonstrate the effectiveness of the proposed algorithm and provide insights about the impact of traffic distribution and content caching on transmit power and backhaul usage of ABSs. 
\end{abstract}


\IEEEpeerreviewmaketitle
\section{Introduction}
\IEEEPARstart{U}{nmanned} aerial vehicles (UAVs), also known as drones or aerial base stations (ABSs), can play an important role in future wireless networks. ABSs have salient attributes; they can make wireless networks more flexible and agile, and act as an additional layer for existing heterogeneous networks (HetNets). ABSs can assist ground BSs by injecting capacity into the network during hard-to-predict congestion times or by temporarily covering users when ground BSs are not accessible, such as in remote areas or in the case of a natural disaster. 

\subsection{Related Work}
There has been a growing interest in the use of ABSs over the past few years \cite{eliVTC, iremmag, 7470933, 8660516, 7470932, 7484835, 8579209}. One of the most challenging issues in integrating ABSs\footnote{What we refer to as ABS in this paper, is a study item for 3GPP Release-17 and is referred to as "on-board radio access node (UxNB)" in the 3GPP documents\cite{UxNB}.} in wireless networks is finding their optimum 3D positions. This was not an issue in terrestrial networks, given that service providers would install BSs in specific areas on the basis of the average traffic in a region. If the average traffic changed over time, the capacity or coverage could be increased/decreased by installing new cells or by switching off existing ones, but once the BSs were installed, their locations were fixed. Because of this, the use of ABSs is very promising as they add a degree of freedom to the network by their adjustable height and their potential mobility, yet finding the best 3D placement for an ABS in relation to terrestrial BSs and other ABSs remains a complex problem.

Several papers have investigated the problem of the 3D placement of ABSs in networks.  Some have focused on finding the optimal trajectory or 3D deployment of a single ABS in a network \cite{IremTWC, AlzenadLetter, 8648498}. In \cite{IremTWC}, the authors assumed that users will move to a location with better coverage if they are offered incentives and they jointly found the 3D location of an ABS and incentives suggested to users. In \cite{AlzenadLetter}, an algorithm was proposed by decoupling the vertical dimension from the horizontal dimension, to find the maximum number of covered users while the transmit power was minimized. For the vertical dimension, the optimum angle that maximized the coverage radius was found, followed by the optimum height. Also, in the horizontal dimension, the deployment was modeled as a circle placement problem. Utilizing solar-powered ABSs is a promising approach to address limited operational time in ABSs. By harvesting solar energy and converting it into electrical energy, longer endurance flights becomes possible; therefore, in \cite{8648498}, the authors considered a solar-powered UAV and suggested an off-line and two online algorithms to jointly find the trajectory of the UAV and resource allocation to serve a set of users while maximizing the system sum throughput during a period of time.

To model more practical scenarios, other papers have considered networks with more than one ABS or with the integration of terrestrial networks as in \cite{8377340, eliVTC, 8610014, 8533634, 7841993, azizi}. 
Machine learning is an important enabler in such involved problems; therefore, in \cite{8377340}, using discounted reward reinforcement learning method, the authors determined the optimum 3D placement of an ABS in a network with ground BSs while user mobility is also taken into account.
In \cite{eliVTC}, a heuristic algorithm based on particle swarm optimization (PSO) was suggested for finding the minimum number of ABSs and their 3D placements in order to meet the minimum quality-of-service (QoS) requirements of the users. In \cite{8610014}, two heuristic approaches, namely genetic algorithm and PSO, were used to maximize the user satisfaction with provided data rates and to find the user-BS associations and optimal positions of ABSs. In \cite{8533634}, an integration of ABSs and drone users was proposed. ABSs were deployed on the basis of the notion of truncated octahedron shapes, and cell associations for minimizing the latency was performed using optimal transport theory. In \cite{7841993}, the trajectory and deployment of ABSs to minimize the power consumption of the users was investigated. In \cite{azizi}, the total transmit powers of the users were minimized while satisfying some QoS constraints in the uplink, and by converting the problems into subproblems, the resource allocations, user associations, and placements of ABSs were found.

Another major aspect of ABSs is their wireless backhaul. Unlike ground BSs, which are usually connected to the core network by high capacity fibre links, ABSs should have a wireless backhaul of limited capacity. Moreover, wireless backhaul capacity may also change dramatically due to inclement weather conditions, environmental parameters, and the distance of an ABS from a backhaul source. Therefore, it is crucial to consider a wireless backhaul with limited variable capacity in designing the network with ABSs. There are a few works that have considered this issue \cite{eliICC, eliPIMRC, 8482444, 9014076, 8755983}. In \cite{eliICC}, the optimum placement of an ABS was found using a network-centric or user-centric approach for maximizing the network throughput. In this work, the backhaul had a fixed capacity and the effect of its variation on the throughput of the network was investigated. In \cite{eliPIMRC}, a number of ABSs were connected to a macro base station (MBS) for backhauling, and the problem jointly optimized the 3D placement of ABSs and the user-BS associations and bandwidth allocations in order to maximize the total rate of the users in a proportional fairness approach.
To overcome the limited backhaul capacity of ABSs, in \cite{8482444}, non-orthogonal multiple access (NOMA) was employed on backhaul transmission, and the authors jointly optimized the resource allocation at the MBS and ABSs along with the decoding order of the NOMA process and the positions of the ABSs to maximize the sum of achievable rate of the users. In \cite{9014076}, the total data rate of the users is maximized in a network with multiple ABSs with fixed altitudes, while taking into account the backhaul capacity limitations of ABSs. In \cite{8755983}, an interference management algorithm is proposed to maximize the sum rate of the network while optimizing the user-BS association and power allocations. In this paper, the dependency between the backhaul and access links is considered in a binary fashion in a way that there will be no transmissions in access side if the received signal to interference plus noise ratio (SINR) at backhaul links are below a predefined threshold. In \cite{9042882}, an integrated network of a satellite, and several MBSs, small BSs, and ABSs are considered where the satellite and MBSs provide backhaul connectivity for small BSs and ABSs. The problem is formulated using a competitive market setting and by applying a heavy ball iterative algorithm, the user-BS association and resource allocation are jointly found and the total profit in the network is maximized.

The diversity of use cases in 5G is classified in three broad categories:
enhanced mobile broadband (eMBB), massive machine-type communications (mMTC), and ultra-reliable and low latency communications (uRLLC) \cite{7894280}. All of them have different challenging demands such as seamless user experience, very low latency or very high data rates. A comprehensive study about the key technologies that should be employed in 5G wireless systems in order to support applications with stringent QoS requirements is provided in \cite{wong_schober_ng_wang_2017}. Edge-caching is one of the propitious techniques to address very low latency requirements of future wireless networks by bringing content closer to end-users via distributed storage in the network. Edge-caching also helps alleviate backhaul congestion especially during peak traffic times in networks with limited backhaul capacity \cite{8119518, 8603721}. Having wireless backhaul in ABSs and using terrestrial BSs as hubs for backhauling can be an issue for some delay-sensitive applications. Moreover, ABSs also have limited battery life. Therefore, it is crucial to equip ABSs with local caches to help with both the backhaul congestion and the latency issue. Enabling ABSs with caching will also reduce energy consumption and prolong battery life. 

To increase the probability of users requesting files that are available in the local cache of an ABS, content caching is often designed using popularity distributions, e.g., the Zipf distribution \cite{Cha2007}. However, files may be cached at an ABS whose channel to the user requesting the file is poor. In this case, data transmission may not be reliable or high transmit power will be required. Therefore, we may have to send the requested files via backhaul to other ABSs who have better channels to that user or directly via the terrestrial BSs in the network. The caching placement phase is usually implemented during an offline interval when the ABS is being charged at an ABS park as illustrated in Fig.~\ref{BHCaching}. When the ABS is on duty, the delivery phase happens after users send their requests. There are a few papers which consider caching in ABSs. In \cite{7875131}, an algorithm based on the machine learning framework of conceptor based echo state networks was proposed to find user-ABS associations, the optimal location of ABSs, and the contents to be cached at each ABS while increasing the quality of experience of users and minimizing the transmit power of ABSs. 
In \cite{8717714}, ABSs captured videos on live games and sent them to small BSs that served the virtual reality users. To meet the delay requirements, BSs were able to cache some popular contents from the data. A distributed deep learning algorithm based on echo liquid state machine was proposed to solve the optimization problem. 
In \cite{8614433}, a network of ABSs that can operate on both licensed and unlicensed bands are considered and an optimization problem is formulated that tries to maximize the number of stable queue users in the network by applying a distributed algorithm based on the machine learning framework of liquid state machine. Using the algorithm, the cloud could predict content request distribution of the users and determine the contents that need to be cached in each ABS. Moreover, each ABS could decide autonomously about its spectrum allocation. 
In \cite{8576651}, throughput among Internet of Things (IoT) devices was maximized by finding the location of the ABS and the placement of content caching. 

To sum-up, utilizing ABSs in wireless networks is very promising, yet a number of issues remain to be effectively addressed, including 3D positioning of ABSs, wireless backhauling, user-BS associations for users with different QoS requirements, and resource allocation.

\subsection{Our Contribution}
As mentioned above, an important difference between a ground BS and an ABS is the major limitation in the backhaul link of ABSs; therefore, considering this constraint in deploying ABSs is essential. To this end, we provide a novel framework that accounts for limited backhaul capacity in ABSs and distinct types of users in the network. In so doing, we propose an algorithm that determines a backhaul-aware 3D placement for ABSs while minimizing their transmit power; further, the algorithm we propose also determines user-BS associations and their bandwidth allocations in a HetNet. In the literature, determining user-BS associations is usually considered in isolation from other aspects of network design; however, for the association of a user with a BS several considerations must be taken into account. These include the channel condition between the BS and the user, the user type, the requested content, and the QoS demand. Therefore, efficient utilization of resources and association of users is coupled with the deployment of ABSs in the network. To the best of our knowledge, this is the first work that considers limited variable backhaul capacity as one of the design constraints in a vertical HetNet and jointly finds 3D location of ABSs, user-BS associations, and bandwidth allocations. The distinctive features of this paper are outlined as follows:

\begin{itemize}
	\item We consider wireless backhaul with limited variable capacity for ABSs. The capacity is affected by the network parameters and the distance from the MBS. This limitation affects both the association of the users and the 3D placement of the ABSs.
	\item One of the main differences in 5G networks and beyond is the capacity to provide seamless service for different applications with diverse demands. Accordingly, we assume two types of users in the network: delay-sensitive and delay-tolerant, and find the user-BS associations and the 3D placement of the ABSs considering user types and their requested contents. 
	\item In contrast to a number of previous studies that considered only ABSs in a network, here we consider a vertical HetNet that includes both ground and aerial BSs. The ground BS can provide service for the users and is also considered as the backhaul source for ABSs.
	\item One of the important add-ons to the ABSs is content caching to alleviate network congestion and decrease latency, especially for delay-sensitive users. We assume that ABSs can cache some popular contents when they are not on-duty. When a user triggers a request, it can be responded to by both ABSs or MBS, depending on the location, requested content, and type of application.
	\item The optimization problem is decoupled into two parts. The 3D placement problem is a non-convex quadratic constraint quadratic programming (QCQP) problem which is transformed into a convex problem using a semi-definite relaxation (SDR) technique, while the association problem is transformed into a binary linear programming and  solved by existing optimization tools.
	\item We cancel the effect of co-channel interference in ABSs by employing ideal directional antennas and show the importance of applying interference mitigation techniques in aerial networks by illustrating the effect of such interference on the performance of the network when directional antennas are not ideal. 
 
\end{itemize}

The rest of this paper is organized as follows. In Section II, the system model is introduced. The problem is described in Section III, and the algorithm is proposed in Section IV. A performance evaluation is presented in Section V, and finally conclusions are drawn in Section VI.

\section{System Model}

\begin{figure}[t]
	\begin{center}
		\includegraphics[width=3.5in]{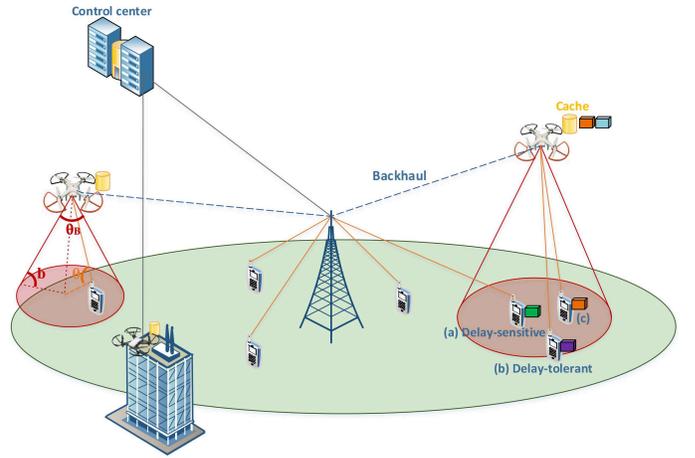}\\
	\end{center}
	\caption{Graphical illustration for the integration of cache-enabled ABSs in a cellular network. Only users that are in LoS coverage of an ABS can be associated with it. User (a) is delay-sensitive and its requested content is not available in the ABS; therefore, it has to be served by the MBS. User (b) is delay-tolerant; therefore, it can connect to the ABS, even though its requested content is not available in that ABS. User (c) requests the data that is cached in the ABS; so whether it is delay-tolerant or delay-sensitive does not matter, and it can be served by the ABS. The ABS park is for energy charging and content caching.}
	\label{BHCaching}
\end{figure}

We consider a downlink wireless HetNet including two tiers of BSs, an MBS, and several ABSs as depicted in Fig.~\ref{BHCaching}. ABSs utilize wireless connections for both access and backhaul links. Wireless links provide a mobility advantage to ABSs such that they can be positioned relative to user locations, which can increase spectral efficiency and decrease average path loss. However, wireless links can be less reliable compared to wired connections, and energy expenditure increases too. Therefore, a careful system design is key for ABS operations. 
We assume that high capacity fibre links carry information from the MBS to the core network; therefore, there is no congestion in the backhaul link of the MBS. We also consider MBS as a hub to connect ABSs to the network. Wireless links are capacity-limited and may suffer congestion during peak network times. They can also increase the latency compared to a wired backhaul; therefore, we assume that ABSs can cache some data. Caching popular content is a very effective method to alleviate backhaul congestion and decrease delays in delivering service to end-users by bringing the data closer to the users. It will also decrease the total transmit power in the network as there is no need to fetch data every time a user requests data from the core network. Hit probability is a principal metric that shows the probability of a user's requested content being stored in a BS or not \cite{8387202}. A lower probability means that more backhaul capacity has to be consumed. Different placement strategies yield different hit probabilities. The mechanism is commonly designed on the basis of the popularity of the content. It is observed that content popularity follows a generalized Zipf law which states that the request rate $q(n)$ for the $n$-th most popular content is proportional to $\frac{1}{n^\alpha}$ for some $\alpha$ \cite{6193511}. Typically, $\alpha$ is between 0.64 and 0.83.  

Let us consider two groups of users in the system, namely delay-tolerant and delay-sensitive users. Such users can be defined either on the basis of the application they use or the fee they pay for their subscribed services. Delay-sensitive users are prone to high latency, while delay-tolerant users can tolerate some delay and receive service at a later time. To overcome the latency issue, a delay-sensitive user should either associate with an MBS that has a wired backhaul to the core network or connect to an ABS that has the requested data in its local cache to avoid the need for a 2 hub connection from the ABS to the core network and consequently lessen the delay.

We consider centralized decision-making in our network whereby a central entity (can be the MBS) is aware of the necessary information and the network parameters and makes the user association decision along with bandwidth allocation, power level transmission, and ABS locations in a region of a network. At the end of section \ref{3DLocation}, a partially distributed decision-making framework is also discussed.

We denote by $\mathcal{I}$ the set of users and by $\mathcal{J}$ the set of BSs. We use $i\in \mathcal{I}=\{1,2,...,I\}$ and $j\in \mathcal{J}=\{0,1,...,J\}$ to index users and BSs, respectively. Index $0$ in $\mathcal{J}$ denotes the only MBS in the system.

\begin{figure}[t]
	\begin{center}
		\includegraphics[width=3in]{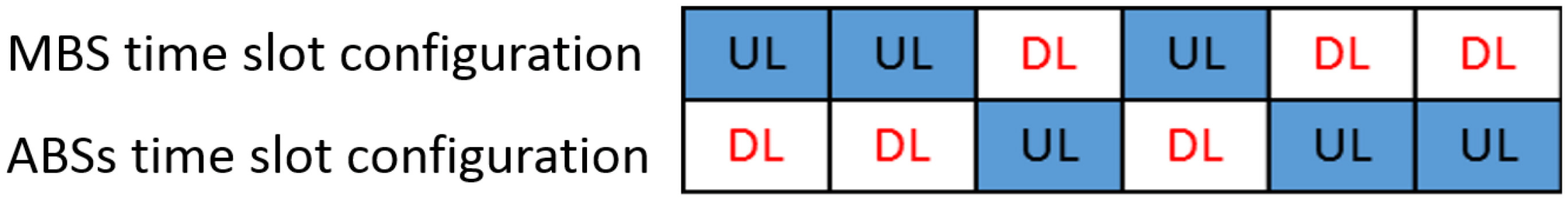}\\
	\end{center}
	\caption{Reverse-TDD time slot configuration for MBS
		and ABSs in a vertical HetNet.}
	\label{slot}
\end{figure}

\begin{figure}[t]
	\begin{center}
		\includegraphics[width=3in]{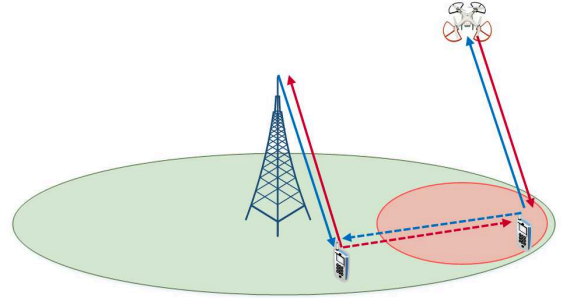}\\
	\end{center}
	\caption{When the MBS is in the DL/UL mode, the ABSs are in the UL/DL mode; hence, the user that is served by a BS in a tier, receives interference from the users in the other tier (dashed arrow). Arrows with similar colors show that they are sharing the same time and frequency slot.}
	\label{interference}
\end{figure}

To avoid interference between backhaul and access links of ABSs, we assume that backhaul links are operated at mmWave frequencies.
We also consider wireless point-to-point Xn links between BSs, which do not interfere with access and backhaul links. Due to non-ideal Xn connections and the energy cost of wireless links, real-time coordination for interference management among BSs may not be efficient.
Hence, to decrease inter-cell interference in different tiers, reverse time division duplex (TDD) is employed, which uses reversed uplink (UL)/downlink (DL) time slot configurations for MBS and ABSs as seen in Fig. \ref{slot}.
When the MBS is in the DL mode, the ABSs are in the UL mode and vice versa. As a result, the only interference the users receive is from the users that are using the same frequency spectrum and associated with a BS in the other tier as observed in Fig. \ref{interference}. Such interference is usually negligible as the channels between users usually have substantially higher path loss compared to channels between BSs and users and the transmit power of the user is usually much less than that of a BS \cite{7386685}. Such interference can also be mitigated by utilizing fast steerable antennas in users devices and directing the signals towards the serving BSs. 


\subsection{Channel Models}

\subsubsection{ABS-to-user}
The path loss between an ABS and a user depends on the altitude of the ABS and its elevation angle from the user. There are mainly two propagation groups corresponding to the receivers with line-of-sight (LoS) connections and those with non-line-of-sight (NLoS) connections which can still receive the signal from the transmitter due to strong reflections and diffractions \cite{7037248}. The total average power reduction of a signal transmitted from an ABS to a ground user can be written as
\begin{equation}\label{fspl}
\textsf{PL (\textrm{dB})} = \textsf{FSPL} +\psi_k,
\end{equation}
where $\textsf{FSPL}$ is the free space path loss according to Friis equation and it is equal to $20 \log(\frac{4\pi f_c d}{c})$, where $f_c$ is the carrier frequency, $c$ stands for the speed of light, and $d$ stands for the distance between the transmitter and the receiver. Also $\psi_k, k=\{\textrm{LoS},\textrm{NLoS}\}$ shows the excess path loss due to the LoS or NLoS channel between the ABS and the user.

The probability of establishing a LoS connection between an ABS and a user can be formulated as
\begin{equation}
P(\textrm{LoS}) = \frac{1}{1+\kappa \exp (-\zeta(\frac{180}{\pi}\theta -\kappa))},
\end{equation}
where $\kappa$ and $\zeta$ are constant values depending on the environment, and $\theta$ is the elevation angle equal to $\arcsin (\frac{z}{d})$, where $z$ is the altitude of an ABS, and $d$ is its distance from a user. Furthermore, the probability of NLoS is $P(\textrm{NLoS}) =1-P(\textrm{LoS})$.

To mitigate the effect of co-channel interference, we assume that ABSs are equipped with directional antennas. The antenna gain can be approximated by \cite{7486987}
\begin{equation}
g = 
\begin{dcases}
g_0, -\frac{\theta_B}{2}\le \phi \le \frac{\theta_B}{2},\\
g(\phi), \text{otherwise},
\end{dcases}
\end{equation}
where $|\phi|=90 - \theta$, $\theta_B$ denotes the ABS directional antenna's half-power beamwidth, and $g_0  \approx \frac{30,000}{\theta_B^2}$ is the maximum gain of the directional antenna \cite{Balanis}. We assume $g(\phi)$ is negligible.

\subsubsection{MBS-to-user}
We adopt the MBS channel model from 3GPP \cite{3GPP1}. The average path loss in dB can be expressed as $15.2 +37.6 \log_{10}(d)$, where $d$ is the distance between the MBS and the user in meters. Moreover, we assume that MBS has omni-directional antenna.

\subsubsection{MBS-to-ABS}
We assume that wireless backhaul links from the MBS to all ABSs experience the LoS propagation condition. The average path loss in dB at 28 GHz is given as $61.4 + 20 \log_{10}(d)$, where $d$ is the distance between the MBS and ABSs in meters \cite{6834753}.

\section{Power Minimization Problem}
In this section, we propose an optimization framework that determines the user-BS associations and bandwidth allocations in addition to the 3D locations of the ABSs to minimize the total transmit power of the ABSs in downlink transmission. First, we introduce the constraints along with the design objectives considered in the system and then present the transmit power minimization problem formulation.

\subsection{System Constraints}
\subsubsection{BS Association Constraints}
In our framework, each user should be served by only one BS. This yields the following constraint:
\begin{equation}
\sum_{j\in \mathcal{J}}\rho_{ij} = 1 ,\;\forall{i\in \mathcal{I}},
\end{equation}
where $\rho_{ij} \in \{0,1\}$ is a binary association indicator variable for user $i$ and BS $j$, and 1 indicates association.

\subsubsection{Bandwidth Allocation Constraints}
The total amount of resources allocated by each BS to all the users cannot exceed the available bandwidth of that BS. Therefore, 
\begin{equation}
\sum_{i \in \mathcal{I}} \rho_{ij} \beta_{ij} \le 1, \;\forall j \in \mathcal{J},
\end{equation}
where $\beta_{ij} \in [0,1]$ is the normalized resource of BS $j$ that is assigned to user $i$.

\subsubsection{Data Rate Constraints}
The wide range of services requested by the users makes their demands fairly disparate. To ensure that data rate demands of the users are met, each user's rate must not be less than its target rate. Therefore,
\begin{equation}
\sum_{j \in \mathcal{J}} \rho_{ij} B  \beta_{ij}  r_{ij} \ge \eta_i, \; \forall i \in \mathcal{I},
\end{equation}
where  $\eta_i$ is the data rate demanded by user $i$, $B$ is the total bandwidth of a BS, variable $r_{ij}$ is the received spectral efficiency of user $i$ when connected to BS $j$, and $\eta_i$ is the minimum required rate of user $i$. Assuming Shannon capacity is achieved, $r_{ij} = \log_2(1+\gamma_{ij})$,
where $\gamma_{ij}$ is the SINR received by user $i$ from $j$th BS.

\subsubsection{User Type Constraints}
We assume that the total data in the network is $K$ files. We also consider a finite cache capacity in each ABS, which means that each ABS can store part of the whole data in its local cache and refresh the contents periodically. Caching generally has a different time scale compared to other parts of the wireless communication systems. Therefore, in this paper we assume that the ABSs have already stored some contents in their local cache using placement algorithms.

Let us define the cache matrix $E^{J\times K}=[e_{jk}]=\{ 0,1 \}$ for ABSs, where $e_{jk}=1$ denotes that ABS $j$ caches $k$th file and $e_{jk}=0$ indicates the opposite. The user request matrix is also defined by $U^{I\times K}=[u_{ik}]=\{ 0,1 \}$, where $u_{ik}=1$ means that user $i$ requests file $k$ and $u_{ik}=0$ means the opposite. We assume that the central entity is aware of both matrices $E$ and $U$ and therefore, can control the caching strategy by obtaining the cache association matrix $F^{I\times J}=[f_{ij}]=\{ 0,1 \}$, where $f_{ij} = 1$ means that the content requested by user $i$ is cached in ABS $j$; otherwise, $f_{ij}=0$. Here is an example to explain the caching strategy in more detail: Assume that there are 10 contents available in the network and each ABS can cache 20 percent of the total contents. Based on a certain placement strategy, ABS 1 decides to keep contents 2 and 4 in its local cache; therefore, $e_{12}=1$ and $e_{14}=1$. On the user side, if user 3 requests for content 4, $u_{34}$ becomes 1. As the central entity is aware of the whole matrix $E$ and $U$, it can obtain the entries of matrix $F$. In the aforementioned example, $f_{31}=1$ as the requested content of user 3 is available in ABS 1.

Let us consider the case of delay-sensitive users only associating with the MBS or ABSs if they have their requested data in their local cache, hence 
\begin{equation}
\sum_{j \in \mathcal{J}} f_{ij} \rho_{ij} \ge \tau_i,\; \forall{i \in \mathcal{I}},
\end{equation}
where $\tau_i \in \{0,1\}$. $\tau_i = 1$ indicates that the user $i$ is delay-sensitive and $\tau_i = 0$ indicates the opposite. We consider $f_{ij}=1$ for $j=0$ as there is a wired connection from the MBS to the core network.

\subsubsection{LoS Constraints for Association}
Adjustable altitudes in ABSs can increase the likelihood of establishing an LoS connection to ground users. A weaker channel implies higher transmit power and resource usage; therefore, to decrease the transmit power, we assume that user $i$ can be associated with ABS $j$ only if it has an LoS channel with a high probability with that ABS. Therefore,
\begin{equation}
P(\textrm{LoS})_{ij} \ge a \rho_{ij},\; \forall i \in \mathcal{I}, \forall j \in \mathcal{J}\backslash 0,
\end{equation}
where $a$ is a number close enough to one.

\subsubsection{Backhaul Capacity Constraints}
The total data rate an ABS can support should not exceed its backhaul capacity. Note that by storing the content in the local cache of ABSs, we can alleviate the backhaul consumption. Accordingly,
\begin{equation}
\sum_{i \in \mathcal{I}} \rho_{ij} B \beta_{ij} r_{ij} (1-f_{ij}) \le C_j,\; \forall{j \in \mathcal{J}\backslash 0}, \label{BH1}
\end{equation}
where $C_j$  is the backhaul capacity of ABS $j$. 

\subsection{Problem Formulation}
Communication technologies are currently responsible for around five percent of the total generated carbon footprint, and this amount is expected to increase significantly with the proliferation of data traffic and the number of connected devices \cite{7446253}. Therefore, due to both economic and environmental concerns, it is not only data rate and throughput that are important factors in network deployment these days, but also how much energy is spent delivering that rate. Moreover, by considering the limited energy in ABSs the importance is doubled. To address the issue of energy efficiency, we optimize the locations of ABSs to minimize the total transmit power (to all the users) considering the data rate requirements of the users and limited backhaul capacity of the ABSs.
Therefore, the optimization problem is expressed as follows:
%

\begin{subequations}
	\begin{alignat}{2}
	\textbf{\textrm{P1:}}& \min_{ \{P_{ij}\},\{\boldsymbol{l}_j \} ,\{ \rho_{ij} \} , \{ \beta_{ij} \} }
	\sum_{j \in \mathcal{J}} \sum_{i \in \mathcal{I}} P_{ij}  \rho_{ij}\\
&\text{subject to}\nonumber\\
&		\sum_{j\in \mathcal{J}}\rho_{ij} = 1 , \;\forall{i\in \mathcal{I}},\label{association}\\
&	\sum_{i \in \mathcal{I}} \rho_{ij} \beta_{ij} \le 1, \; \forall j \in \mathcal{J},\label{max-BW}\\
& \sum_{j \in \mathcal{J}} \rho_{ij}  B  \beta_{ij}  r_{ij}  \ge \eta_i, \; \forall i \in \mathcal{I},\label{min-rate}\\
& \sum_{j \in \mathcal{J}} f_{ij}  \rho_{ij} \ge \tau_i, \; \forall{i \in \mathcal{I}},\label{delay}\\
&P(\textrm{LoS})_{ij} \ge a \rho_{ij}, \forall i \in \mathcal{I},\forall j \in \mathcal{J}\backslash 0, \label{LoS}\\
&\sum_{i \in \mathcal{I}} \rho_{ij}   B  \beta_{ij} r_{ij} (1-f_{ij}) \le C_j, \; \forall j \in \mathcal{J}\backslash 0,\label{BH}\\
&	\rho_{ij} \in \{0,1\},
	\beta_{ij} \in [0,1], \forall i \in \mathcal{I}, \; \forall j \in \mathcal{J},
	\end{alignat}
\end{subequations}
where $P_{ij}$ is the transmit power from BS $j$ to user $i$ and $\boldsymbol{l}_j=(x_j,y_j,z_j)$ is the 3D location of ABS $j$. This optimization problem has a non-convex objective function and nonlinear constraints with a combination of binary and continuous variables. In other words, it is a non-convex, NP-hard optimization problem.

\section{Proposed Algorithm}

To alleviate the difficulties mentioned in the preceding section, we break the problem down into subproblems and solve them iteratively until they converge into a local optimum. First, for fixed $\{ \rho_{ij} \}$ and $\{ \beta_{ij} \}$, we find $\{ \boldsymbol{l}_j \}$, and then for new 3D locations of ABSs, we update $\{ \rho_{ij} \}$ and $\{ \beta_{ij} \}$. Finally $P_{ij}$ is calculated using the updated information. The detail is explained in the following.

\subsection{3D Locations of ABSs} \label{3DLocation}
In this step, we assume that $\{ \rho_{ij} \}$ and $\{ \beta_{ij} \}$ are known and hence we find $\{\boldsymbol{l}_j\}$.

According to \eqref{association}, each user is associated with only one BS; therefore, \eqref{min-rate} can be simplified to
\begin{equation}
\rho_{ij} B  \beta_{ij} r_{ij} \ge \eta_i\rho_{ij},\;\forall {i\in \mathcal{I}}, \forall j \in \mathcal{J}. 
\end{equation}

By replacing $r_{ij}$, we have
\begin{equation}
\label{snr}
\log_2(1+\gamma_{ij})\rho_{ij}\ge \frac{\eta_i}{B\beta_{ij}}\rho_{ij},
\end{equation}
where $\gamma_{ij}= \frac{P_{ij}g_{ij}}{h_{ij}N_0B\beta_{ij}}$. Variable $h_{ij}$ is the path loss between BS $j$ and user $i$, $g_{ij}$ is the antenna gain, and $N_0$ denotes the noise power spectral density of the additive white Gaussian noise (AWGN). If a user is associated with an ABS, $h_{ij}=(\frac{4\pi f_c}{c})^2 10^ {\psi_k/10} d_{ij}^2$, where $d_{ij}$ is the distance between ABS $j$ and user $i$. By substituting $\gamma_{ij}$ in \eqref{snr} and after some manipulations we have
\begin{equation}
P_{ij}\rho_{ij} \ge  g_{ij}^{-1}h_{ij} (2^{\frac{\eta_i}{B\beta_{ij}}}-1)N_0B\beta_{ij}\rho_{ij},
\end{equation}
which can be simplified as 
\begin{equation}
P_{ij} \rho_{ij} \ge A_{ij} d_{ij}^2 \rho_{ij},
\end{equation}
where $A_{ij}=g_{ij}^{-1}(\frac{4\pi f_c}{c})^2 10^ {\psi_k/10}  (2^{\frac{\eta_i}{B\beta_{ij}}}-1) N_0 B \beta_{ij}$ and  $d_{ij}=((x_j-x_i)^2+(y_j-y_i)^2+z_j^2)^{1/2}$, where $(x_j,y_j,z_j)$ and $(x_i,y_i)$ are the coordinates related to the location of ABS $j$ and user $i$, respectively. Therefore, in $\textbf{\textrm{P1}}$ instead of minimizing the transmit power of ABSs, one can minimize the weighted distance between the ABSs and their associated users. 


Inequality \eqref{LoS} can be rewritten as 
\begin{equation}
\theta_{ij} \ge b\rho_{ij}, \forall i \in \mathcal{I}, \; \forall j \in \mathcal{J}\backslash 0, \label{simple_theta}
\end{equation}
where $b=\frac{-\pi}{180 \zeta}\ln \frac{1-a }{\kappa a }+\frac{\pi \kappa}{180}$. Therefore,
\begin{equation}
((x_j-x_i)^2+(y_j-y_i)^2 +V z_j^2) \rho_{ij} \le 0,
\end{equation}
where $V= 1-\frac{1}{\sin ^2 b}$.

Minimizing the total transmit power implies that each user receives service with the minimum required rate; therefore, \eqref{BH} can be changed to
\begin{equation}
\sum_{i \in \mathcal{I}} \eta_i  \rho_{ij} (1-f_{ij}) \le C_j, \; \forall{j \in \mathcal{J}\backslash 0}.
\end{equation}
By considering equal bandwidth allocations between ABSs for backhauling, $C_j=\frac{W}{J} \cdot \log_2(1+\frac{P_0}{h_{j0}N_0 W})$, where $W$ is the total available bandwidth for backhaul connection from an MBS to ABSs, $P_0$ is the MBS transmit power for backhauling, and $h_{j0}$ is the path loss from the MBS to ABS $j$ and it is equal to $10^{6.14}d_{j0}^2$.
After some manipulations, we have 
\begin{equation}
L_j  [(x_j-x_0)^2+(y_j-y_0)^2+z_j^2] \le D,
\end{equation}
where $L_j=2 ^{\frac{J}{W}\sum_{i \in \mathcal{I}} \eta_i \rho_{ij}  (1-f_{ij})} -1$ and $D=\frac{P_0}{10^{6.14} N_0 W }$.

Finally, $\textbf{\textrm{P1}}$ is simplified to
\begin{subequations}
	\label{3Dlocation}
	\begin{alignat}{2}
	\textbf{\textrm{P2:}} &\!\min_{ \{{x_j, y_j , z_j} \}} 
	 \sum_{j \in \mathcal{J}\backslash 0} \sum_{i \in \mathcal{I}} A_{ij} [(x_j-x_i)^2+(y_j-y_i)^2+z_j^2]  \rho_{ij}\\
	&\text{subject to}\nonumber\\ 	
	&[(x_j-x_i)^2+(y_j-y_i)^2+ V z_j^2]  \rho_{ij} \le 0, \;\forall i\in \mathcal{I} , \forall j \in \mathcal{J}\backslash 0, \label{los-constraint}\\
	&	L_j [(x_j-x_0)^2+(y_j-y_0)^2+z_j^2] - D \le 0,\; \forall j \in \mathcal{J}\backslash 0\label{capacity-constraint}.
	\end{alignat}
\end{subequations}

The optimization problem $\textbf{\textrm{P2}}$ is in the form of separable QCQP problems \cite{5447068} whose general form is given as
\begin{subequations}
	\begin{alignat}{2}
	&\!\min_{\{\boldsymbol{\xi}_j\}}  \sum_{j \in \mathcal{J}\backslash 0} \frac{1}{2}\boldsymbol{\xi}_j^T\boldsymbol{G_0}_j\xi_j+\boldsymbol{q_0}_j^T\boldsymbol{\xi}_j+n_{0j} \label{QCQP}\\
	&\text{subject to}\nonumber\\
	&\frac{1}{2}\boldsymbol{\xi}_j^T\boldsymbol{G_t}_j\boldsymbol{\xi}_j+\boldsymbol{q_t}_j^T \boldsymbol{\xi}_j + n_{tj} \le 0, \;\forall t=\{1,...,I+1\}, \forall j \in \mathcal{J}\backslash 0.
	\end{alignat}
\end{subequations}

Here we have
\begin{equation}
\boldsymbol{\xi}_j=
\begin{bmatrix}
x_j & y_j &z_j
\end{bmatrix}^T,
\end{equation}
and
\begin{equation}
\boldsymbol{G_0}_j=
\begin{bmatrix}
2 \sum_i A_{ij}\rho_{ij} & 0 & 0\\
0 & 2 \sum_i A_{ij}\rho_{ij} & 0\\
0 & 0 & 2 \sum_i A_{ij} \rho_{ij} 
\end{bmatrix}.
\end{equation}
Also
\begin{equation}
\boldsymbol{q_0}_j =
\begin{bmatrix}
\sum_i -2x_i A_{ij }\rho_{ij} & \sum_i -2y_i A_{ij} \rho_{ij} & 0
\end{bmatrix},
\end{equation}
and $n_{0j}=\sum_{i}(x_i^2+y_i^2)A_{ij} \rho_{ij}$.

For the constraints we have
\begin{equation}
\boldsymbol{G_t}_j=
\begin{dcases}
\begin{bmatrix}
2 \rho_{ij} & 0 & 0\\
0 & 2 \rho_{ij} & 0\\
0 & 0 & 2V \rho_{ij}
\end{bmatrix}& , \text{if } t=i,\\
\begin{bmatrix}
2L_j & 0 & 0\\
0 & 2L_j & 0\\
0 & 0 & 2L_j
\end{bmatrix} &,            \text{if } t=I+1.\\
\end{dcases}
\end{equation}
Also
\begin{align}
\boldsymbol{q_t}_j =
\begin{dcases}
\begin{bmatrix}
-2x_i \rho_{ij} & -2y_i \rho_{ij} & 0
\end{bmatrix} &, \text{if } t=i,\\
\begin{bmatrix}
-2 L_j x_0 &-2 L_j y_0 & 0 
\end{bmatrix}  &, \text{if } t=I+1,\\
\end{dcases}
\end{align}
and 
\begin{equation}
n_{tj}= 
\begin{dcases}
(x_i^2+y_i^2)\rho_{ij} &, \text{if } t=i,\\
L_jx_0^2+L_jy_0^2-D &, \text{if } t=I+1.
\end{dcases}
\end{equation}

Here we note that $\boldsymbol{G_t}_j$ is not a positive semidefinite (PSD) matrix; therefore, the QCQP problem is not convex. To solve this issue, we apply the $\textit{Suggest-and-Improve}$ framework to get approximate solutions to the non-convex QCQP problem \cite{boydd}. 

\subsubsection{Suggest}
In this step a $\textit{candidate point}$ is found. To find a candidate point, we transform the problem to a semi-definite programming problem using the SDR technique. SDR is a computationally efficient approximation approach to QCQP.

Using $\boldsymbol{\xi}_j^T\boldsymbol{G_0}_j\xi_j = \Tr(\boldsymbol{G_0}_j\boldsymbol{\xi}_j\boldsymbol{\xi}_j^T)$ and by introducing the new variable $\boldsymbol{X}_j = \boldsymbol{\xi}_j \boldsymbol{\xi}_j^T$ we can rewrite each QCQP problem for every ABS as
\begin{subequations}\label{qcqp-transform}
	\begin{alignat}{2}
	&\!\min_{\{\boldsymbol{X}_j\},\{\boldsymbol{\xi}_j\}}~~~
	\frac{1}{2} \Tr(\boldsymbol{G_0}_j \boldsymbol{X}_j) +\boldsymbol{q_0}_j^T\boldsymbol{\xi}_j+n_{0j}\\
	&\text{subject to} \nonumber\\
	&\frac{1}{2}\Tr(\boldsymbol{G_t}_j \boldsymbol{X}_j)+\boldsymbol{q_t}_j^T \boldsymbol{\xi}_j + n_{tj} \le 0, \;\forall t=\{1,...,I+1\}, \\
	&  
	\boldsymbol{X}_j = \boldsymbol{\xi}_j \boldsymbol{\xi}_j^T. \label{non-convex}
	\end{alignat}
\end{subequations}

Using this transformation, we have embedded the original problem with three variables into a much larger space by obtaining the additional property that the objective and constraints are affine in $\boldsymbol{X}_j$ and $\boldsymbol{\xi}_j$ except the constraint \eqref{non-convex} which is not convex. Problem \eqref{qcqp-transform} can be relaxed into a convex problem by replacing this non-convex equality constraint with a PSD constraint $\boldsymbol{X}_j- \boldsymbol{\xi}_j \boldsymbol{\xi}_j^T \succeq 0$. Then by solving the following convex problem in each ABS, a lower bound on the optimum value of the problem $\textbf{\textrm{P2}}$ is found.

\begin{subequations}
	\label{SDR}
	\begin{alignat}{2}
	&\!\min_{\{\boldsymbol{X}_j\},\{\boldsymbol{\xi}_j\}}  ~~~
	\frac{1}{2} \Tr(\boldsymbol{G_0}_j \boldsymbol{X}_j) +\boldsymbol{q_0}_j^T\boldsymbol{\xi}_j+n_{0j}\\
	&\text{subject to} \nonumber\\
	&\frac{1}{2}\Tr(\boldsymbol{G_t}_j \boldsymbol{X}_j)+\boldsymbol{q_t}_j^T \boldsymbol{\xi}_j + n_{tj} \le 0, \;\forall t=\{1,...,I+1\}, \\
	& 
	\begin{bmatrix}
	\boldsymbol{X}_j & \boldsymbol{\xi}_j \\
	\boldsymbol{\xi}_j^T  & 1
	\end{bmatrix} \succeq 0,
	\end{alignat}
\end{subequations}
where the last constraint is formulated as a Schur complement \cite{Boyd}. This problem is convex and can be conveniently solved by available software packages such as convex optimization toolbox CVX \cite{cvx} in MATLAB.

Afterwards, one can apply the randomization procedure explained in Algorithm \ref{GRP} to improve the solution found via SDR. Gaussian randomization procedure minimizes the expected objective function subject to holding the constraints in expectation; therefore, there is no guarantee that the sampling points from the normal distribution give feasible points of problem $\textbf{\textrm{P2}}$ at all. However, these points are a good choice for the $\textit{Suggest}$ method and can serve as a starting point for the $\textit{Improve}$ method.

\subsubsection{Improve}
In this step, we run a local method to find a solution point that is not worse than the candidate point. As the candidate points might be infeasible, the goal in this step is to minimize the constraint violation and obtain a smaller value in the objective function. Here we apply a $\textit{coordinate-descent}$ method to improve the candidate point $\boldsymbol{\delta}_{l^*j}$. Coordinate-descent method is an algorithm that solves optimization problems by successively performing minimizations along coordinate directions to find the local minimum of a function, and it includes two phases. 

The goal in the first phase is to reduce the maximum constraint violation and achieve a feasible point if possible. Let us consider $\boldsymbol{\delta}_{l^*j}$ as the candidate point. We repeatedly cycle over each coordinate of $\boldsymbol{\delta}_{l^*j}$ and update it to the value that minimizes the maximum constraint violation. If the violations on all constraints along all the coordinates become zero or smaller, a feasible point $\boldsymbol{\gamma}_{j}$ is found.

In the second phase, we look for other feasible points with better objective values. To do so, we cycle over each coordinate of $\boldsymbol{\gamma}_{j}$ and optimize the objective function while all the constraints are satisfied. In other words, we solve $\textbf{\textrm{P2}}$ with all the variables fixed but one.

As shown above, the problem of finding the 3D placements of the ABSs was transformed to separable QCQPs; therefore, for decision making, a partially distributed solution is possible where the 3D location of each ABS is computed locally at that particular ABS, but the decisions for associations are made at a central controller. The distributed solution is indeed preferred due to not relying on a single controller in the network. Because if the centralized system suffers from a failure, the operation of the entire network is disrupted. A centralized solution may also endure signaling overhead, outdated information, and scalability problems \cite{7206589}. 

\begin{algorithm}[tb]
	\small
	\caption{Gaussian randomization procedure for the QCQP problem}
	\label{GRP}
	\begin{algorithmic}[1]
		\State \textbf{Inputs:} SDR solution $\boldsymbol{X}_j^*$ and $\boldsymbol{\xi}_j^*$, the number of randomizations $L$.
		\For {l = 1, ..., L}
		\State generate $\boldsymbol{\delta}_{lj} \sim \mathcal{N} (\boldsymbol{\xi}_j^*,\boldsymbol{X}_j^*-\boldsymbol{\xi}_j^* \boldsymbol{\xi}_j^{*T})$.
		\EndFor
		\State determine $l^*=\argmin_{l=1,...,L} \frac{1}{2}\boldsymbol{\delta}_{lj}^T\boldsymbol{G_0}_j\boldsymbol{\delta}_{lj}+\boldsymbol{q_0}_j^T\boldsymbol{\delta}_{lj}+n_{0j}$.
		\State \textbf{output:} $\boldsymbol{\delta}_{l^*j}$, which is the approximate solution for the QCQP problem.
	\end{algorithmic}
\end{algorithm}

\subsection{User-BS Associations and Bandwidth Allocations}
In the second step, we assume that the locations of ABSs are known and hence the problem is transformed to 
\begin{subequations}
	\begin{alignat}{2}
	\textbf{\textrm{P3:}} &\!\min_{ \{ \rho_{ij} \} , \{ \beta_{ij} \}} ~~~
	\sum_{i \in \mathcal{I}} \sum_{j \in \mathcal{J}} g_{ij}^{-1}h_{ij} (2^{\frac{\eta_i}{B_j\beta_{ij}}}-1)N_0B\beta_{ij}\rho_{ij}\\
	&\text{subject to} \nonumber\\
	&\sum_{j\in \mathcal{J}}\rho_{ij} = 1 ,\;\forall{i\in \mathcal{I}},\\
	&  
	\sum_{i \in \mathcal{I}} \rho_{ij}  \beta_{ij} \le 1, \;\forall j \in \mathcal{J},\\
	&   
	\sum_{j \in \mathcal{J}} f_{ij}  \rho_{ij} \ge \tau_i, \;\forall{i \in \mathcal{I}},\\
	&    
	\rho_{ij} \le P(\textrm{LoS})_{ij}/\alpha, \;\forall{i\in \mathcal{I}}, \forall{j \in \mathcal{J}\backslash 0}, \\
	&            
	\sum_{i \in \mathcal{I}} \eta_i  \rho_{ij}  (1-f_{ij}) \le C_j, \;\forall{j \in \mathcal{J}\backslash 0},\\
	&          
	\rho_{ij} \in \{0,1\},
	\beta_{ij} \in [0,1], \forall{i\in \mathcal{I}}, \forall{j \in \mathcal{J}}.
	\end{alignat}
\end{subequations}

This problem is non-convex due to the non-convexity of the objective function, binary variables $\{ \rho_{ij} \}$, and the product relationship between $\{ \rho_{ij} \}$ and $\{ \beta_{ij} \}$. To circumvent this difficulty, we consider equal resource allocation between all the users that are associated with a BS and assume all the users in LoS coverage of an ABS are associated with that ABS; therefore, $\textbf{\textrm{P3}}$ is transformed to
\begin{subequations}
	\label{user-association}
	\begin{alignat}{2}
	&\!\min_{ \{ \rho_{ij} \}} ~~~
    \sum_{i \in \mathcal{I}} \sum_{j \in \mathcal{J}} g_{ij}^{-1}h_{ij} (2^{\frac{\eta_i}{B_j\beta_{ij}}}-1)N_0B\beta_{ij}\rho_{ij}\\
	&\text{subject to} \nonumber\\
	&\sum_{j\in \mathcal{J}}\rho_{ij} = 1 ,\;\forall{i\in \mathcal{I}},\\
	& 
	\sum_{j \in \mathcal{J}} f_{ij} \rho_{ij} \ge \tau_i, \;\forall{i\in \mathcal{I}},\\
	& 
	\rho_{ij} \le P(\textrm{LoS})_{ij}/\alpha, \;\forall{i\in \mathcal{I}}, \forall{j \in \mathcal{J}\backslash 0},  \\
	& 
	\sum_{i \in \mathcal{I}} \eta_i \rho_{ij} (1-f_{ij}) \le C_j, \;\forall{j \in \mathcal{J}\backslash 0},\\
	&
	\rho_{ij} \in \{0,1\}, \forall{i\in \mathcal{I}}, \forall{j \in \mathcal{J}}.
	\end{alignat}
\end{subequations}

Since this problem is a binary linear programming problem, it can be solved by optimization tools such as MOSEK \cite{mosek}. 
To refine $\{\beta_{ij}\}$ values, we solve the problem below in each ABS $j$ for known $\{\rho_{ij}\}$.
\begin{subequations}\label{bw}
	\begin{alignat}{2}
	&\!\min_{ \{ \beta_{ij} \}} 
	~~~\sum_{i \in \mathcal{I}} 
	g_{ij}^{-1}h_{ij} (2^{\frac{\eta_i}{B_j\beta_{ij}}}-1)N_0B\beta_{ij}\rho_{ij} \label{obj-bw}\\
	&\text{subject to} \nonumber\\
	&\sum_{i \in \mathcal{I}} \rho_{ij}  \beta_{ij} \le 1,\\
	&    
	\beta_{ij} \in [0,1], \forall{i \in \mathcal{I}}.
	\end{alignat}
\end{subequations}
Optimization problem \eqref{bw} is convex as the objective function and all the constraints are convex. The convexity of \eqref{obj-bw} is proved by showing that its second derivative is greater than zero for all positive values of $\{\beta_{ij}\}$; therefore, \eqref{bw} can be solved efficiently by existing optimization tools. Algorithm \ref{main_alg} explains the iterative procedure that solves the two subproblems and updates the variables until convergence to a local optimum. It should be noted that although the main problem \textbf{P1} is always feasible, by dividing it to subproblems and iteratively solving them, we may face infeasible solution for subproblem \textbf{P2}. This is due to the fact that $\{\rho_{ij}\}$ found in subproblem \textbf{P3}, can force an ABS to move to a higher altitude based on \eqref{los-constraint} to provide a good LoS channel, while at the same time it can force the ABS to move to a lower altitude and closer to MBS to satisfy capacity constraint in \eqref{capacity-constraint}, and therefore, there might exist no feasible solution for this subproblem. 

\begin{algorithm}[tb]
	    \small
	\caption{Finds 3D locations of ABSs, user-BS associations, and bandwidth allocations}
	\label{main_alg}
	\begin{algorithmic}[1]
		\State \textbf{Inputs:} User locations, number of ABSs.
		\State \textbf{Initialization:} Choose initial locations of the ABSs randomly. Set $t=1$, $P(t) = \sum_{i \in \mathcal{I}} \sum_{j \in \mathcal{J}} g_{ij}^{-1}h_{ij} (2^{\frac{\eta_i}{B_j\beta_{ij}}}-1)N_0B\beta_{ij}\rho_{ij}$, and  
		$m(t)= P(t)-P(t-1)$. Define $m(1)=N$, where $N$ is a big number. $\epsilon$ is a small positive number. 
		\While  {$m(t) \ge \epsilon$ }
		\State Find $\rho_{ij}(t)$ and $\beta_{ij}(t)$ in $\textbf{\textrm{P3}}$ and update $P(t)$.
		\State $t = t+1$.
		\State Find $\boldsymbol{l}_{j}$ in $\textbf{\textrm{P2}}$ and update the ABSs locations. Find $P(t)$ and $m(t)$, accordingly.    
		\EndWhile
	\end{algorithmic}
\end{algorithm}

\subsection{Implementation}
In this section we provide an explanation regarding the implementation of our proposed algorithm and investigate the required message passing between different parts of the network.
By applying the proposed method, finding the user-BS associations along with the bandwidth allocations and ABSs locations is an iterative process. The signaling between the radio network and the users is done by radio resource control (RRC) protocol which is mainly responsible for broadcasting the system information and establishment and configuration of radio bearers. All the information related to the channels and topology of the network will be communicated by the central entity through control channels. By knowing the location of the users and the ABSs, the central entity can find the elevation angle between them and calculate the probability of LoS and path loss, accordingly. In the first iteration, after randomly initializing the locations of ABSs, $\boldsymbol{l}_j$, the central entity can find the user-BS associations, $\{\rho_{ij}\}$, by solving subproblem \eqref{user-association}. Then the bandwidth allocations, $\{\beta_{ij}\}$, can be improved by solving subproblem \eqref{bw} and ABSs locations are updated using subproblem \eqref{3Dlocation}. In each iteration all the parameters are updated. These calculations are iteratively done in the central entity without having to physically move the ABSs to their locations. After convergence, the optimum values of $\{\boldsymbol{l}_j\}$, $\{\rho_{ij}\}$, and $\{\beta_{ij}\}$ are obtained and broadcast to the ABSs through Xn interface. Each BS will capture its required information and move to its optimum location; then, they can find their precise channels with the users and update the bandwidth allocations and their locations by solving subproblems \eqref{bw} and \eqref{3Dlocation} to minimize the total transmit power. The information related to the new locations of ABSs and the required bandwidth per user are gathered by the MBS through Xn interface to update the user-BS associations. This information will then be broadcast to the ABSs which can adjust their locations if necessary. 

\subsection{Computational Complexity Analysis}
Problem \textbf{P1} is a non-convex NP-hard problem where finding the optimum solution using exhaustive search is computationally prohibitive. In contrast with the exhaustive search, the iterative algorithm proposed herein, has a polynomial time complexity and therefore, is suitable for solving large scale problems. To analyze the computational complexity, we first derive the complexity of solving \textbf{P2}. 
Convex problem \eqref{SDR} can be solved with a complexity of $\mathcal{O}((I+1)^3)\approx \mathcal{O}(I^3)$ in each ABS, by exploiting the problem structure and building customized interior-point algorithms \cite{5447068}.
To find a better solution for \textbf{P2}, Gaussian randomization procedure is applied which is computationally efficient since the complexity of evaluating the objective function corresponding to the $L$ random samples is  $\mathcal{O}(n^2L)$, where $n$ is the number of variables equal to $3$ \cite{8038863}. 
Also, the coordinate-descent method which is applied to improve the suggested solution for location of ABSs has a computational complexity of $\mathcal{O}(L_{\text{iter}})$, where $L_{\text{iter}}$ is the number of iterations required to find the solution. 
In fact, the complexities of the randomization procedure and coordinate-descent methods are almost negligible compared with that of solving the SDP problem \eqref{SDR}.

Problem \textbf{P3} is divided to 2 subproblems \eqref{user-association} and \eqref{bw}, where both can be efficiently solved by interior-point methods. Such methods are iterative algorithms where each iteration has a cubic complexity and the number of iterations for a given accuracy is at most $\mathcal{O}(m^{0.5})$, where $m$ is the number of constraints. Therefore, $\mathcal{O}((J+1)^{3.5}I^{3.5})$ iterations are required in the worst case to obtain an optimal feasible solution for \eqref{user-association} and the complexity order of solving \eqref{bw} is $\mathcal{O}((J+1)I^{3.5})$, where $J+1$ is the number of BSs in the network. Finally, the number of required iterations to converge to a local optimum solution for \textbf{P1} should also be considered.
Based on this analysis, it is observed that solving \eqref{user-association} which has to be done in a central entity has the highest computational complexity between all the subproblems.

\section{Performance Evaluation}
\begin{figure}[t!]
	\begin{center}
		\subfloat[]{\label{PowerVsUserCoV}\includegraphics[width=2.9in]{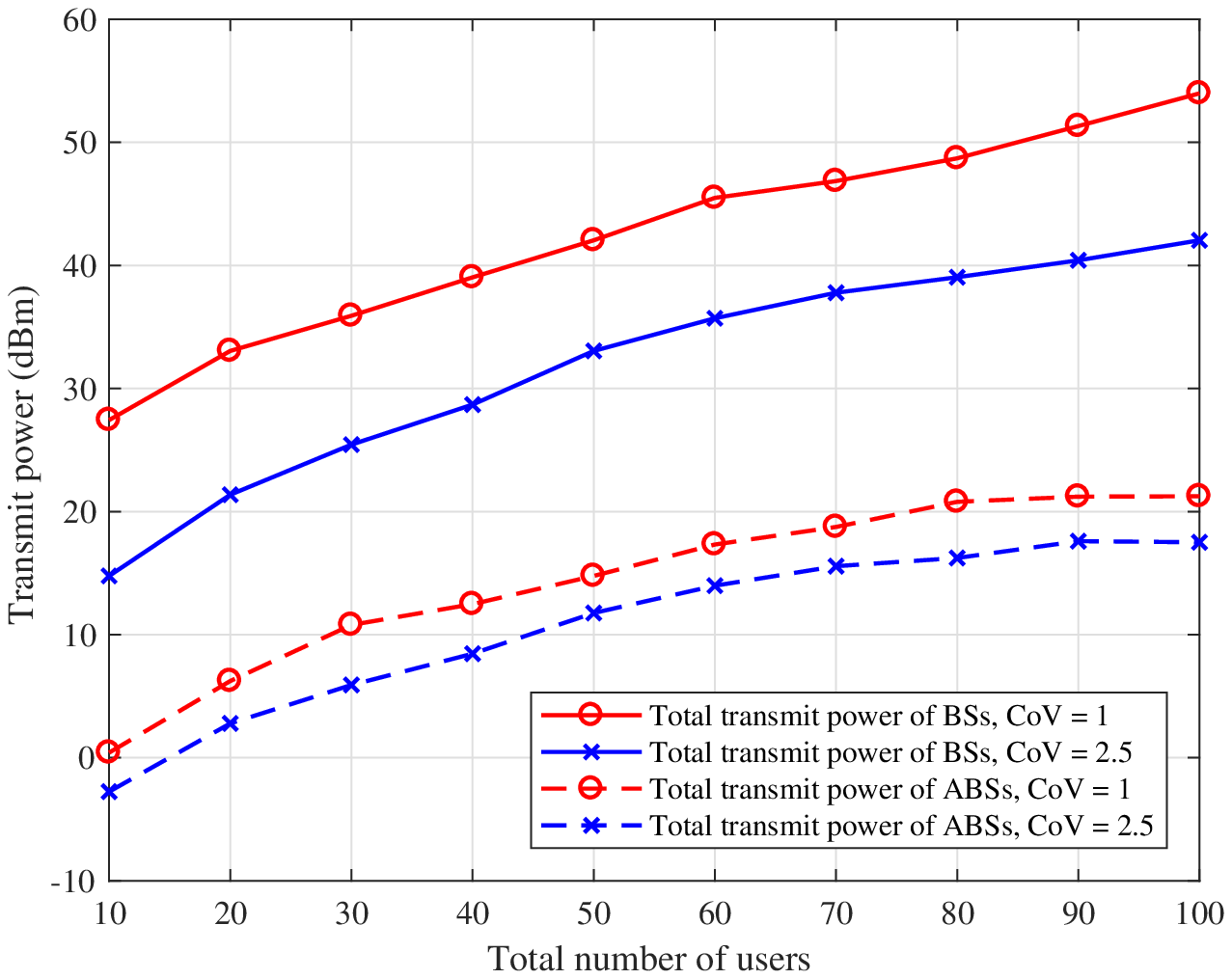}}	
			
		\subfloat[]{\label{ABSuserVsUserCoV}\includegraphics[width=2.9in]{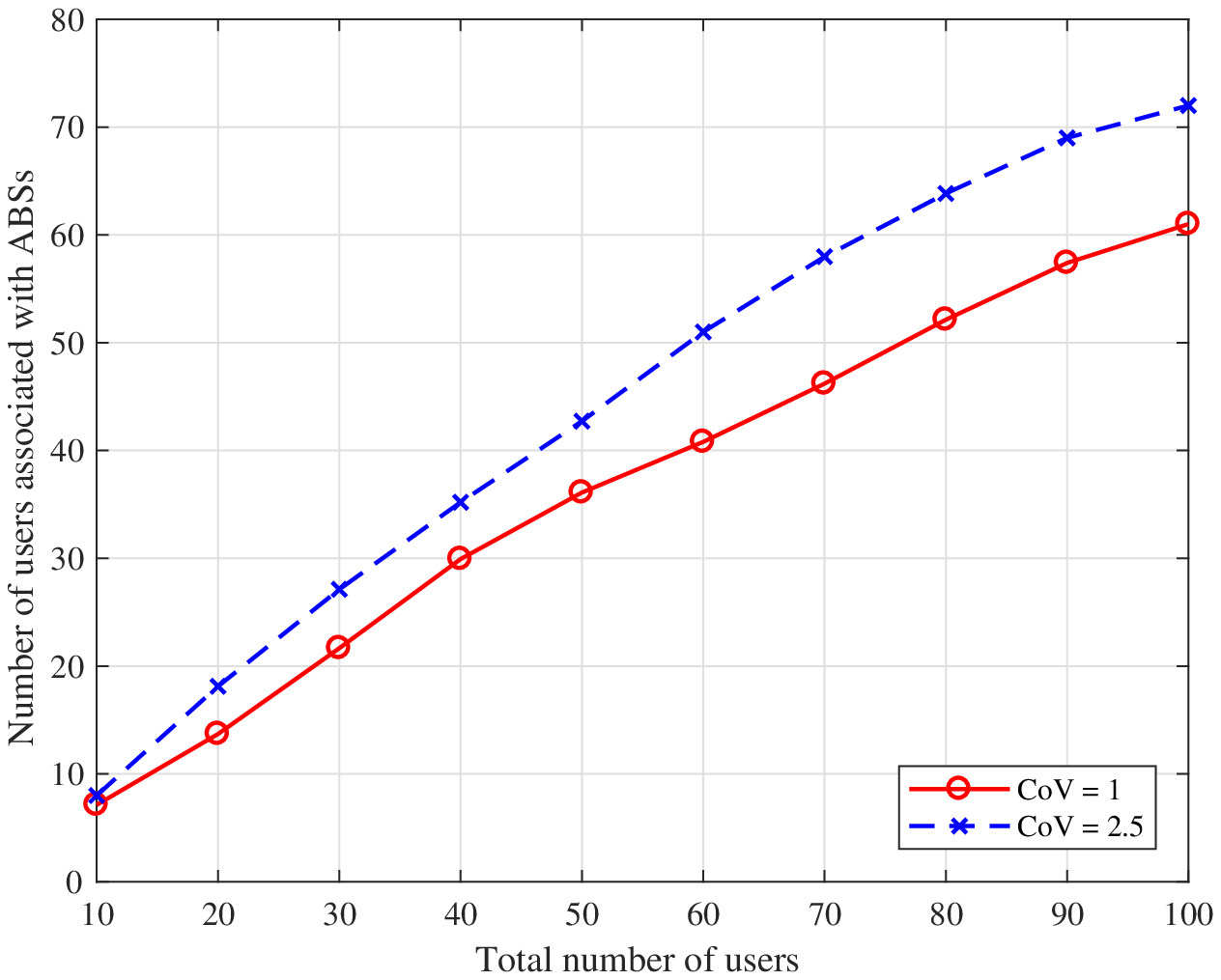}}
	\end{center}
	\caption{Total required transmit power for 2 different CoVs. When the users are more clustered, the required transmit power in ABSs is decreased, although the number of users associated with the ABSs is increased. There are 3 ABSs in the network and each one can cache 20\% of the total contents. Also 10\% of the users are delay-sensitive.}
	\label{UserCoV}
\end{figure}

In this section, we investigate the performance of the proposed algorithm and provide numerical examples to illustrate the merits of these approaches for different network instances. We consider an urban square region with a side length of $1000$ meters and a two-tier HetNet scenario. One MBS is located in the center of the area which can serve both the users and the ABSs (as a backhaul hub). There are also several ABSs in the network whose locations should be determined. We also assume that $\frac{\theta_B}{2}$ is equal to $90-b$; therefore, all the users that are in LoS coverage of an ABS, are also in the main lobe coverage of the directional antennas of ABSs. The bandwidth for each BS in access side is 40 MHz and total bandwidth for backhaul of ABSs is 400 MHz. We assume that the total size of the available data in the system is 10 files and each ABS can store a fixed number of contents in its local cache based on the popularity. For simplicity, we assume that all files have the same size.
\begin{table}[t]
	\centering
	\caption{Simulation Parameters}\label{table1}
	\vspace*{-1.2\baselineskip}
	\begin{tabular}[t]{| c | c | c | c |} 
		\hline
		\textbf{Parameter} & \textbf{Value} & 
		\textbf{Parameter} & \textbf{Value} \\ 
		\hline
		$P_0$ & 40 dBm & $(\kappa,\zeta)$ & (9.61, 0.16) \\  
		\hline
		$\alpha$ & 0.9 &  $(\psi_{\text{LoS}},\psi_{\text{NLoS}})$ & (1 dB, 20 dB) \\
		\hline
		$f_c$ & 2 GHz  & $N_0$ & -170 dBm/Hz  \\
		\hline
		$z_{\text{max}}$& 600 m & noise figure & 10\\
		\hline
	\end{tabular}
\end{table}\noindent
The users are placed according to a uniform or $Mat\acute{e}rn$ distribution in different scenarios where the latter is a doubly Poisson cluster process \cite{martin-haenggi}. In such a distribution, the \textit{parent} points (the centers of the clusters) are created by a homogeneous Poisson process and the \textit{daughter} points (users in our model), are uniformly scattered in circles with a specific radius around \textit{parent} points using another homogeneous spatial Poisson process. We also consider that the user types, i.e., delay-tolerant or delay-sensitive, are known to the BSs and the core network. The users have different data rate demands drawn from a set of $\{$5, 7, 10 Mbps$\}$. Other simulation parameters are provided in \tablename~\ref{table1}. All results are averaged over 100 Monte Carlo simulations.
A typical user distribution with CoV=2 along with an MBS and 2 or 3 ABSs are demonstrated in Fig.~\ref{3D}. An explanation about CoV is provided in the Appendix. As seen, the location of ABSs is determined based on different parameters such as the traffic distribution, the data rate requirements of the users, and the number of ABSs, to name a few. When the number of ABSs is 2 as in Fig.~\ref{3D2ABS}, to increase the chance of establishing an LoS connection with more users, the ABSs move to higher altitudes and consequently they have to increase their transmit power. On the other hand, when the number of ABSs is 3 as in Fig.~\ref{3D3ABS}, the ABSs can move to lower altitudes when the users are highly clustered  in a small area to decrease the required transmit power. As observed, 2 ABSs can serve the same number of users for this typical user distribution simply by changing their 3D locations and their transmit power. Therefore, based on the traffic distribution, user requirements, and power budgets, sometimes using fewer number of ABSs is preferred as they can do the same job with lower complexity in the system. 
\begin{figure}[t!]
	\centering
	\subfloat[]{\label{3D2ABS}\includegraphics[width=3.6in]{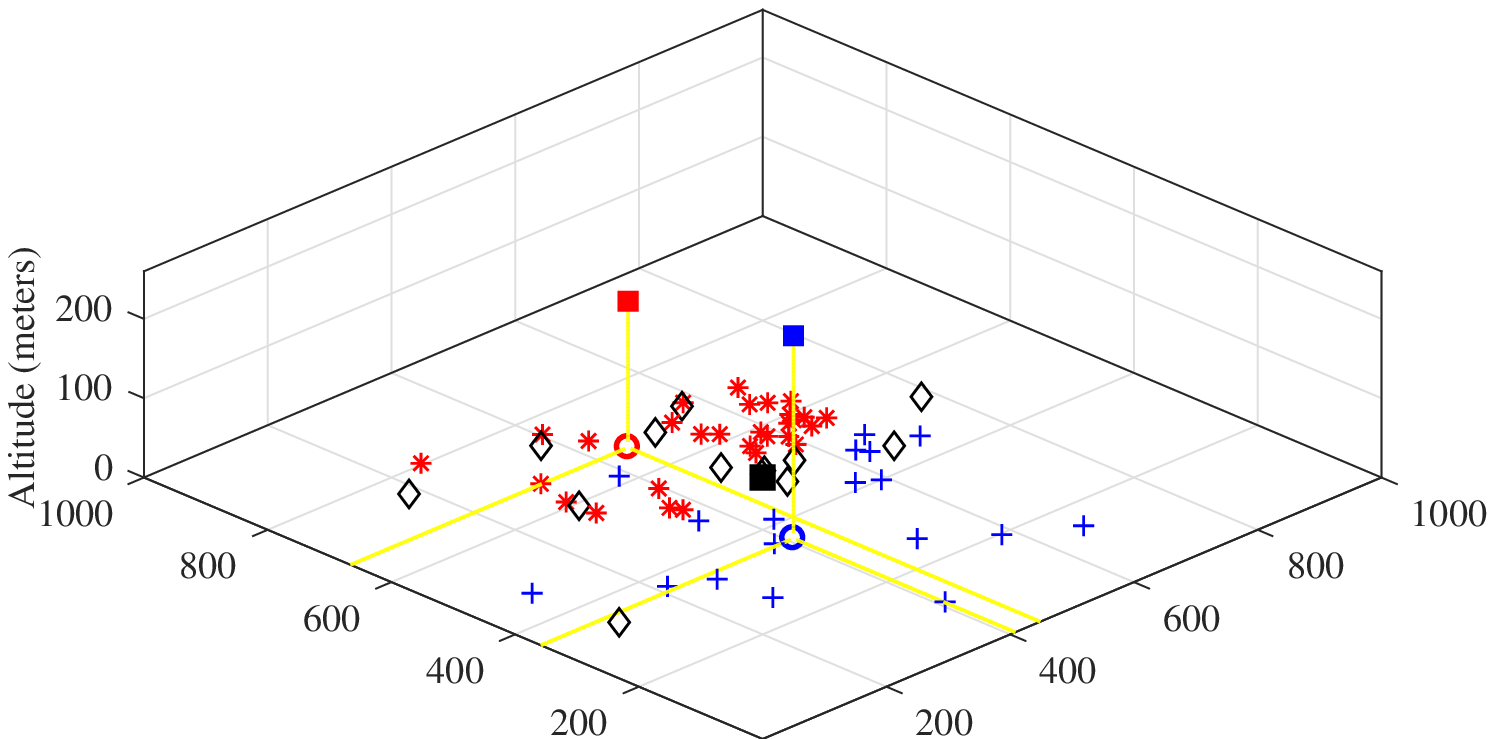}}
	
	\subfloat[]{\label{3D3ABS}\includegraphics[width=3.6in]{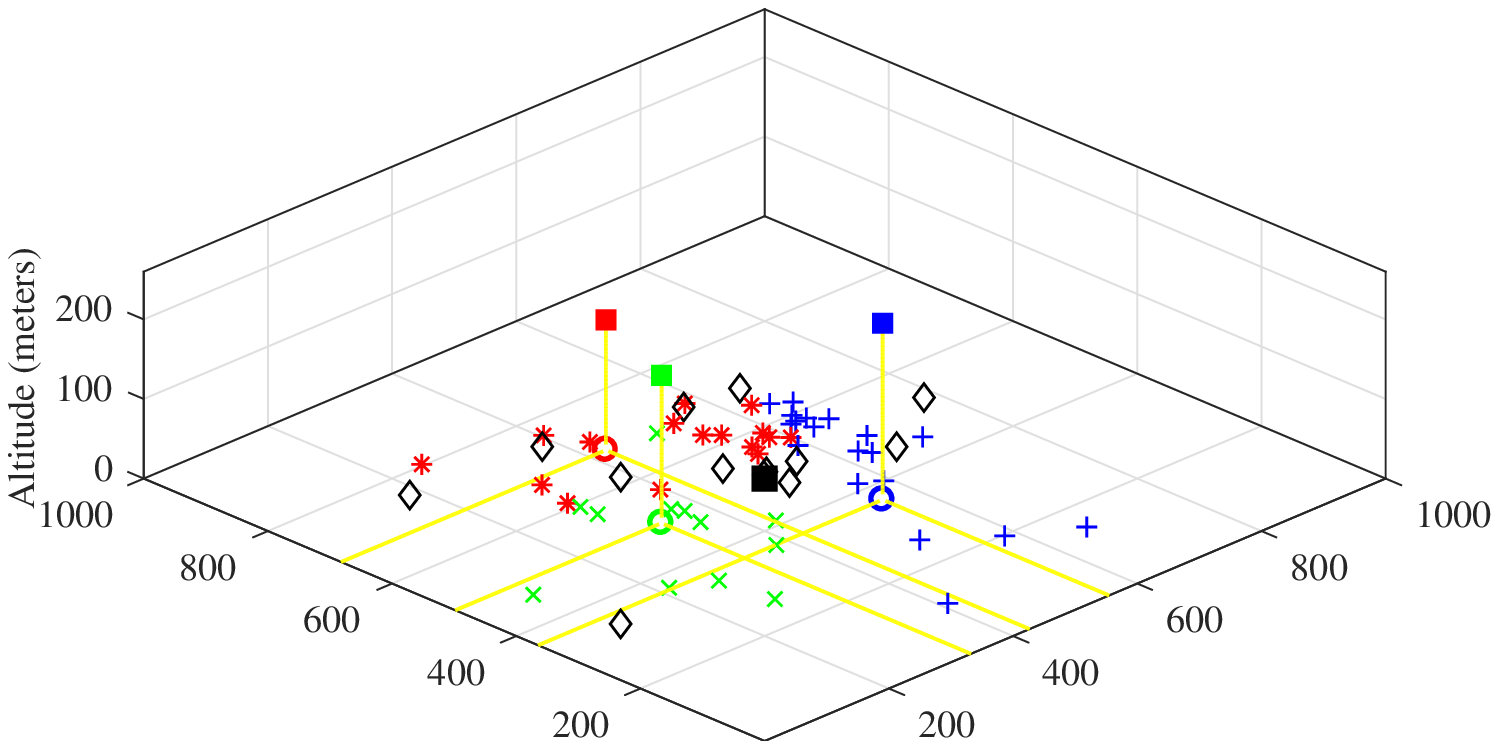}}
	\caption{A typical user distribution with CoV=2 along with the MBS and 3D placement of ABSs: (a) 2 ABSs, (b) 3 ABSs. The MBS is shown in a black square at the center of the coordinate system. ABSs and their projections on the XY-plane are represented by squares and circles, respectively. Also, users and their associated BSs are represented with similar colours. 10\% of the users are delay-sensitive and each ABS can cache 20\% of the total contents.}
	\label{3D}
\end{figure}
In Fig.~\ref{PowerVsUserCoV}, the total transmit power of all the BSs and only ABSs for 2 different CoVs are shown. It is observed that by increasing the CoV, although more users are served by ABSs, as seen in Fig.~\ref{ABSuserVsUserCoV}, the transmit power is decreased. This is due to the fact that by increasing the CoV, the average distance between users and ABSs is reduced, which also decreases the required transmit power. 

\begin{figure}[t!]
	\begin{center}
		\includegraphics[width=2.9in]{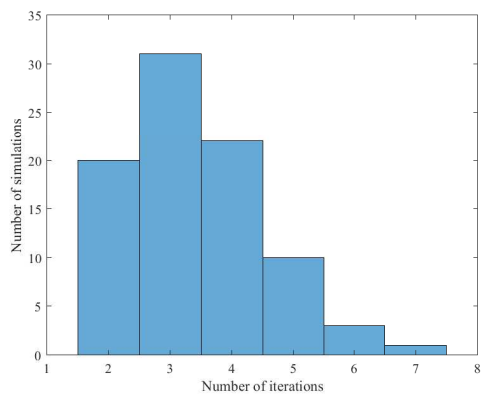}\\
	\end{center}
	\caption{The number of iterations required for convergence to a local optimum. There are 70 users in the network with CoV=1 and 10\% of them are delay-sensitive. Also the number of ABSs is 3, and each ABS can cache 20\% of the total contents.}
	\label{hist}
\end{figure}

Fig.~\ref{hist} shows the number of required iterations to converge to a local optimum across different snapshots. There are 70 users in the network with uniform distribution and 10$\%$ of the users are delay-sensitive. The number of ABSs is 3. As shown, the algorithm converges quickly. In the majority of snapshots the local optimum is found in less than 3 iterations, and the maximum number of iterations in all the snapshots is less than 8. 

\begin{figure}[t!]
	\begin{center}
		\subfloat[]{\label{BHvsUser-caching}\includegraphics[width=2.9in]{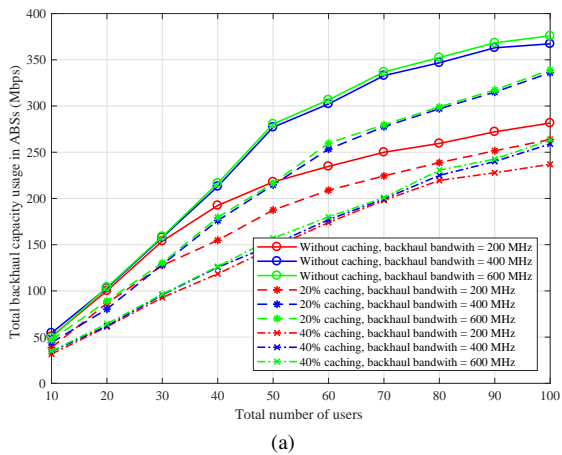}}
				
		\subfloat[]{\label{ABSUservsUser-caching}\includegraphics[width=2.9in]{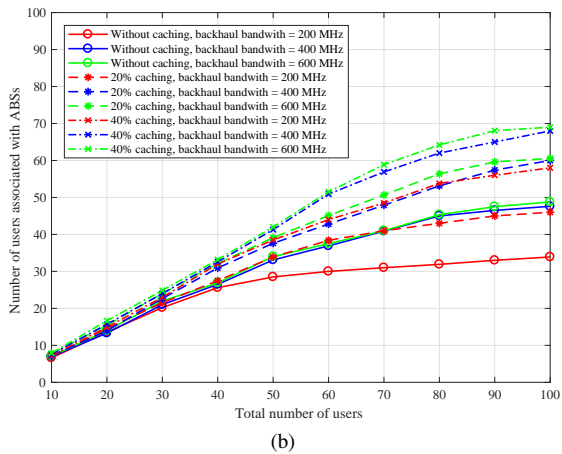}}
	\end{center}
	\caption{(a) Average backhaul capacity usage in each ABS, and (b) average number of users associated with ABSs, with and without content caching in ABSs for different backhaul bandwidths. The number of ABSs is 3 and users are placed according to a uniform distribution with 10\% of them being delay-sensitive.}
	\label{caching}
\end{figure}

Fig.~\ref{caching} depicts the effect of caching in ABSs on backhaul capacity usage and number of users associated with ABSs. As shown in Fig.~\ref{BHvsUser-caching}, the backhaul is less consumed when the ABSs are cache-enabled and by increasing the size of local cache in ABSs, although more users are served by ABSs as observed in Fig.~\ref{ABSUservsUser-caching}, the backhaul capacity usage is reduced. The figure also shows the effect of bachhaul capacity limitation on the number of served users. When the total backhaul bandwidth is increased from 200 MHz to 400 MHz, due to higher capacity, more users are served by ABSs; while by increasing the backhaul bandwidth from 400 MHz to 600 MHz, we almost obtain similar results. This is due to the fact that by increasing the bandwidth, the noise power is also increased and therefore, there will be no huge increment in backhaul capacity by increasing the bandwidth from 400 MHz to 600 MHz. It is also observed that the number of associated users with ABSs are almost the same for different sizes of caching in ABSs when the total number of users is low. By increasing the number of users, the backhaul capacity usage is increased and finally no more users can be associated with ABSs due to limitation in backhaul capacity. This issue happens much faster in ABSs that are not enabled with caching and have low backhaul bandwidth. 

\begin{figure}[t!]
	\begin{center}
		\includegraphics[width=3in]{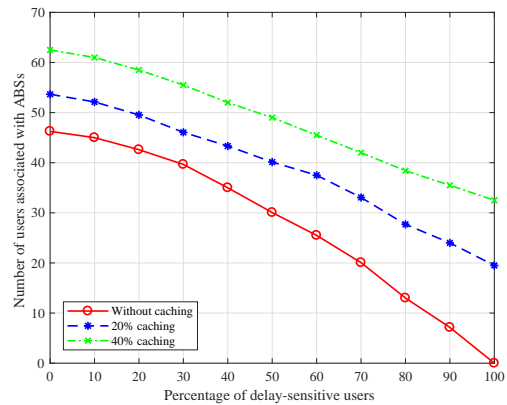}\\
	\end{center}
	\caption{Number of users associated with 3 ABSs for different percentage of delay-sensitive users. The number of users is 80 and they are placed according to a uniform distribution.}
	\label{UserVsDS-caching}
\end{figure}

Fig.~\ref{UserVsDS-caching} illustrates the number of users associated with ABSs for different percentage of delay-sensitive users. By increasing the number of delay-sensitive users, more users have to associate with the MBS, because delay-sensitive users can associate with ABSs only  if their requested content is available in the local cache of an ABS that is in their LoS coverage. Therefore, by increasing the caching size as shown in Fig.~\ref{UserVsDS-caching}, this issue can be alleviated.

\begin{figure}[t!]
	\begin{center}
		\includegraphics[width=3in]{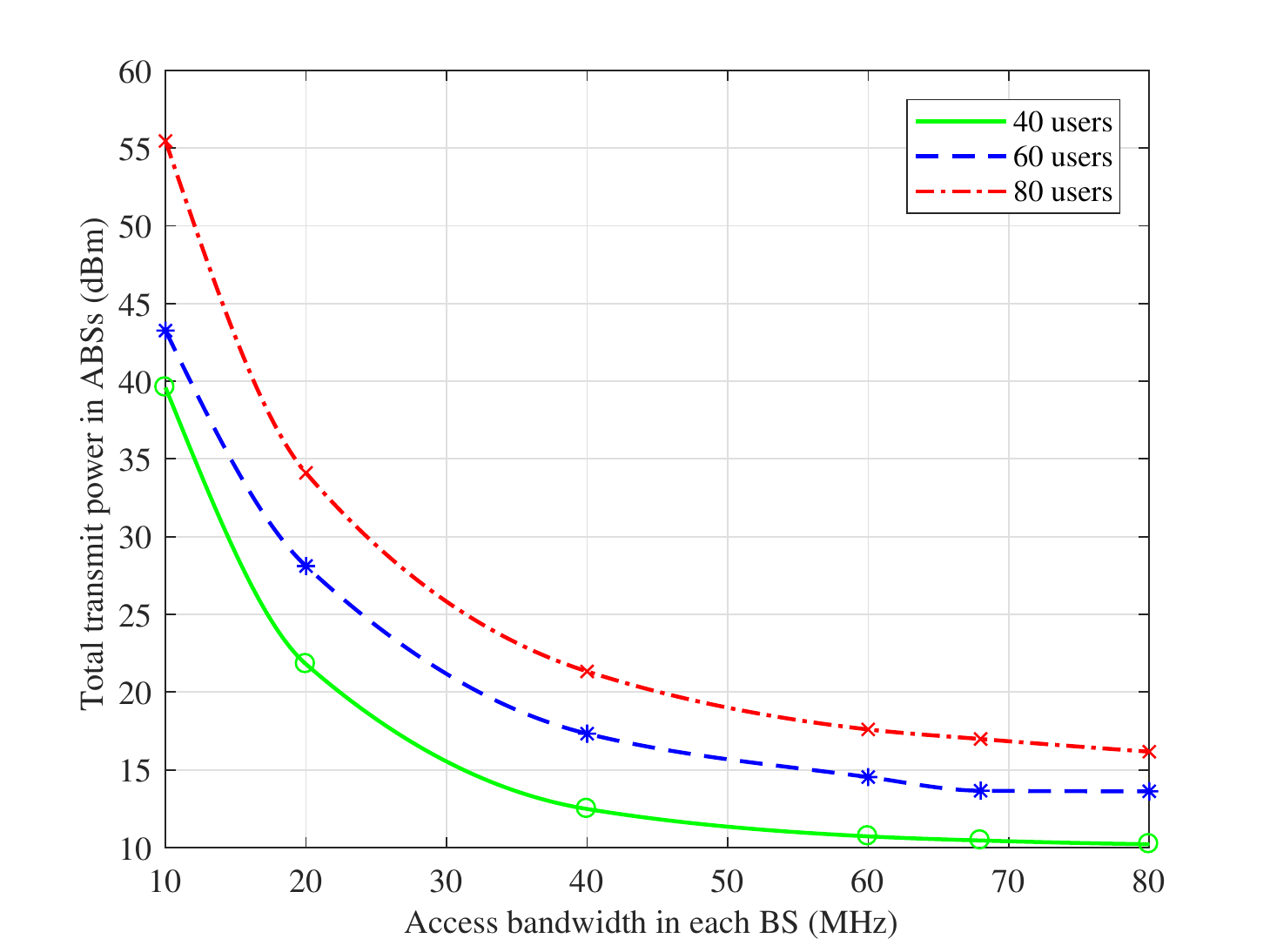}\\
	\end{center}
	\caption{Total transmit power in ABSs for different access bandwidths. There are 3 ABSs in the network and the users are placed according to a uniform distribution (CoV=1). Each ABS can cache 20\% of the content and 10\% of the users are delay-sensitive.}
	\label{PowerVsBW}
\end{figure}

Fig.~\ref{PowerVsBW} shows the total transmit power in ABSs versus the available access bandwidth in BSs for different number of users in the network. As shown, by increasing the available bandwidth, the required transmit power is decreased with a higher rate at first and then the decrement gradually reduces. This is due to the fact that by increasing the bandwidth, the noise power is also increased. It is also observed that by increasing the traffic in the network, the required power and/or bandwidth will increase. Therefore, due to scarcity of available resources $~$in$~$ the network, careful design is essential to meet the requirements of the system, while minimizing the resource consumption in the network.

\begin{figure}[t!]
	\begin{center}
		\subfloat[]{\label{CDFrate}\includegraphics[width=3in]{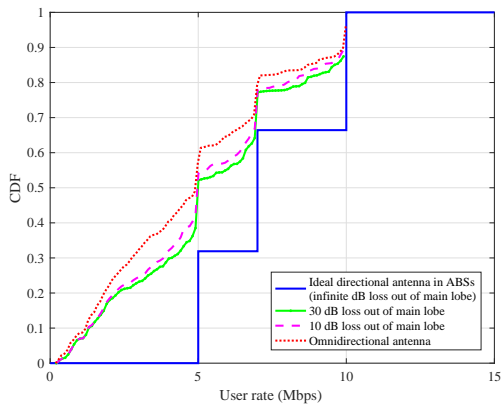}}

		\subfloat[]{\label{CDFrateDtoD}\includegraphics[width=3in]{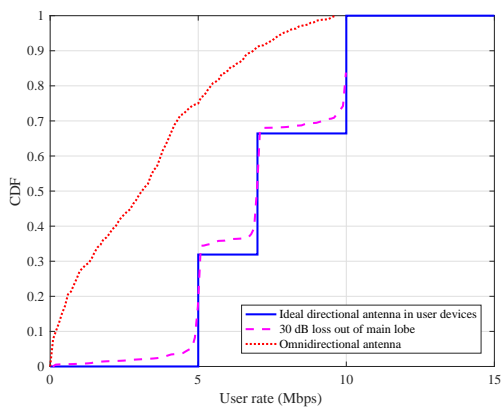}}
	\end{center}
	\caption{(a) CDF of achieved user rates for different antenna gains out of the main lobe in ABSs, (b) CDF of achieved user rates when there is interference from users in the other tier.}
	\label{CDFratekol}
\end{figure}

Fig.~\ref{CDFratekol} shows the effect of interference on the achieved rate of the users.
In our system, we considered ABSs with directional antennas and assumed that the gain out of the main lobe is negligible; however, this is not the case in practice and therefore, co-channel interference exists in the system. Fig.~\ref{CDFrate} illustrates the effect of such interference (in the worst case) on the achieved users' rates. The blue lines show the CDF of target rates; while other curves illustrate the CDF of achieved rates when the side lobe gains are not negligible. As observed, co-channel interference can dramatically decrease the performance of the network; therefore, it is important to mitigate the effect of such interference in radio resource allocations. To do so, in a centralized approach joint optimization among different cells is required; while alternatively, in a decentralized approach, it is possible to achieve similar performance by exchanging messages among ABSs and letting each ABS make its local decision based on the information received from the other ABSs. A promising approach to decrease the interference in ABSs is utilizing multiple antennas to employ interference-aware beamforming. Highly directional beams can provide spatial orthogonalization which enable ABSs deployment with low interference \cite{wong_schober_ng_wang_2017}.
We also investigate the effect of interference caused by users in different tiers on the achieved rates in Fig.~\ref{CDFrateDtoD}. In our system, we assumed user devices with ideal directional antennas with narrow beams towards the serving BS, so there is no interference from other users; however, to see the effect of such interference in practice, we consider two scenarios where the user devices have directional antennas with side lobes or omni-directional antennas. We assume that the transmit power in user devices is 0.1 watts and the average path loss in dB is given as $28 + 40 \log_{10}(d)$, where $d$ is the distance between the users \cite{3GPP1}. As observed, by utilizing steerable antennas in user devices and directing the signals towards the serving BSs, the interference caused by the users is significantly reduced.
\begin{figure}[t!]
	\begin{center}
		\includegraphics[width=3.3in]{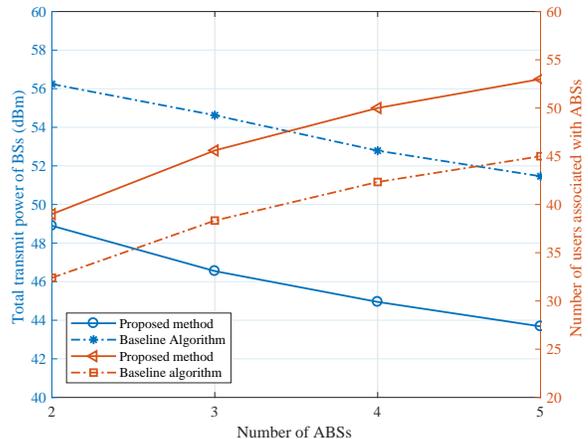}\\
	\end{center}
	\caption{Comparison between our proposed method and a baseline algorithm. There are 70 users in the network with CoV=1 and 10\% of them are delay-sensitive. Also each ABS can cache 20\% of the total contents.} 
	\label{benchmark}
\end{figure}

To show the effectiveness of our proposed algorithm, Fig.~\ref{benchmark} compares the performance of our proposed method with a baseline scheme described in Algorithm \ref{BMA}. As observed in this figure, the performance of the proposed method in terms of the required transmit power and the number of users that are associated with ABSs, is substantially better than the baseline algorithm.

\begin{algorithm}[tb]
	\footnotesize
	\caption{Baseline algorithm}
	\label{BMA}
	\begin{algorithmic}[1]
	    \State \textbf{Inputs:} user locations, number of ABSs ($J$), maximum allowed altitude for ABS ($z_\text{max})$.
		\State Divide users to different groups by k-means clustering. Assume that the number of clusters is equal to the number of ABSs.
		\State Consider the center of the clusters as the projection of ABSs locations on the ground.
		\State Find $\{\rho_{ij}\}$ based on closest horizontal distance between users and ABSs.
		\State Find the altitude of ABS $j$ as $z_j = \text{min} (\text{max}\{z_{ij}\},z_\text{max} )$, where $z_{ij}= d_{ij}\sin b, \forall i  \text{ that } \rho_{ij}=1$.
		\For{$\forall i \in \mathcal{I}$ and $\forall j \in \mathcal{J}$}
		\State \textbf{If} $\rho_{ij}=1$ and $z_{ij} > z_j$ \textbf{then}
		\State $\rho_{ij}=0$ and $\rho_{i0}=1$
		\EndFor
		\For {$\forall j \in \mathcal{J}$}
		\While {$\sum_{i \in \mathcal{I}} \rho_{ij}r_i(1-f_{ij})>C_j$}
		\State based on central entity decision for some $i$ that $\rho_{ij}=1$, put $\rho_{ij}=0$ and $\rho_{i0}=1$
		\EndWhile
		\State Find new $z_j$ based on updated $\rho_{ij}$ variables.
		\State Find allocated bandwidth for each user: $\beta_{ij} = \frac{1}{\sum_{i \in \mathcal{I}} \rho_{ij}}$.
		\EndFor
		\State Find $P = \sum_{i \in \mathcal{I}} \sum_{j \in \mathcal{J}} g_{ij}^{-1}h_{ij} (2^{\frac{\eta_i}{B_j\beta_{ij}}}-1)N_0B\beta_{ij}\rho_{ij}$.		
	\end{algorithmic}
\end{algorithm} 

\section{Conclusion}
In this study, we developed a novel framework to jointly optimize the 3D placement of ABSs, the association of users with BSs, and bandwidth allocations, while minimizing the total transmit power of the BSs. To decrease both the latency and congestion issue in the backhaul, we proposed content caching in the ABSs. Based on different applications in future wireless systems, we defined two groups of users: delay-sensitive and delay-tolerant. It was shown that delay sensitive users could either associate with the MBS or the  ABSs which have their requested content in their local caches, while delay-tolerant users could connect to all the ABSs with LoS coverage or the MBS. Due to the intractability of the problem, we divided the optimization problem into subproblems and iteratively updated them. First, user-BS associations and bandwidth allocations were found using existing optimization tools, and then 3D placements of ABSs were updated using the SDR approach and coordinate-descent method. Simulation results showed that the proposed algorithm yields significant performance gains and indicated that caching can extensively decrease backhaul usage and help congestion issue. It was also shown that in networks with highly clustered users, there is a higher chance of establishing LoS connections between ABSs and users and hence more users can be associated with ABSs. 
\vspace{-0.1cm}
\section{Appendix}
\subsection{Coefficient of Variation (CoV)}
\begin{figure}[t!]
	\centering	\subfloat[]{\label{CoV1}\includegraphics[width=1.2in]{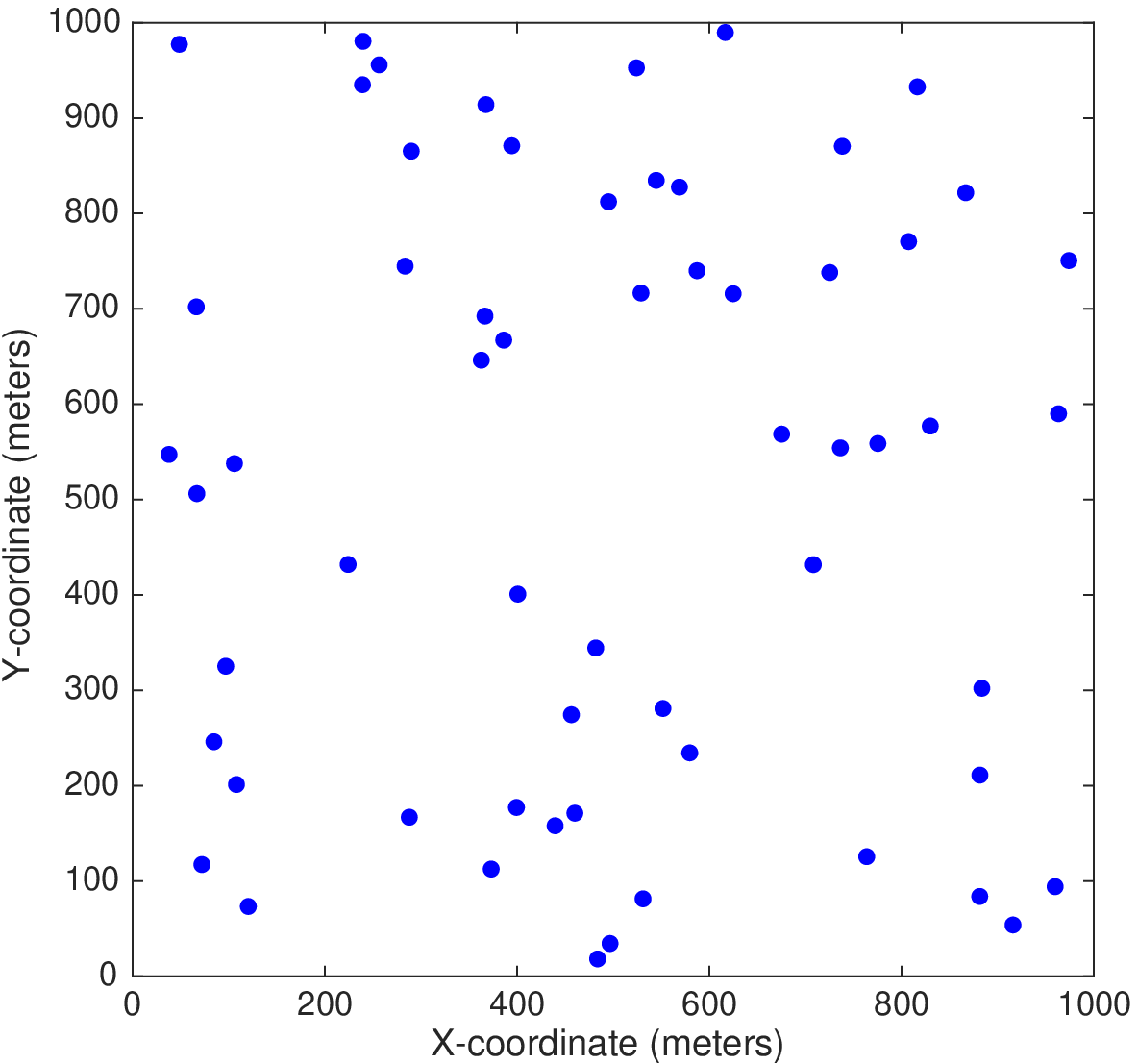}}	
	\subfloat[]{\label{CoV2}\includegraphics[width=1.06in]{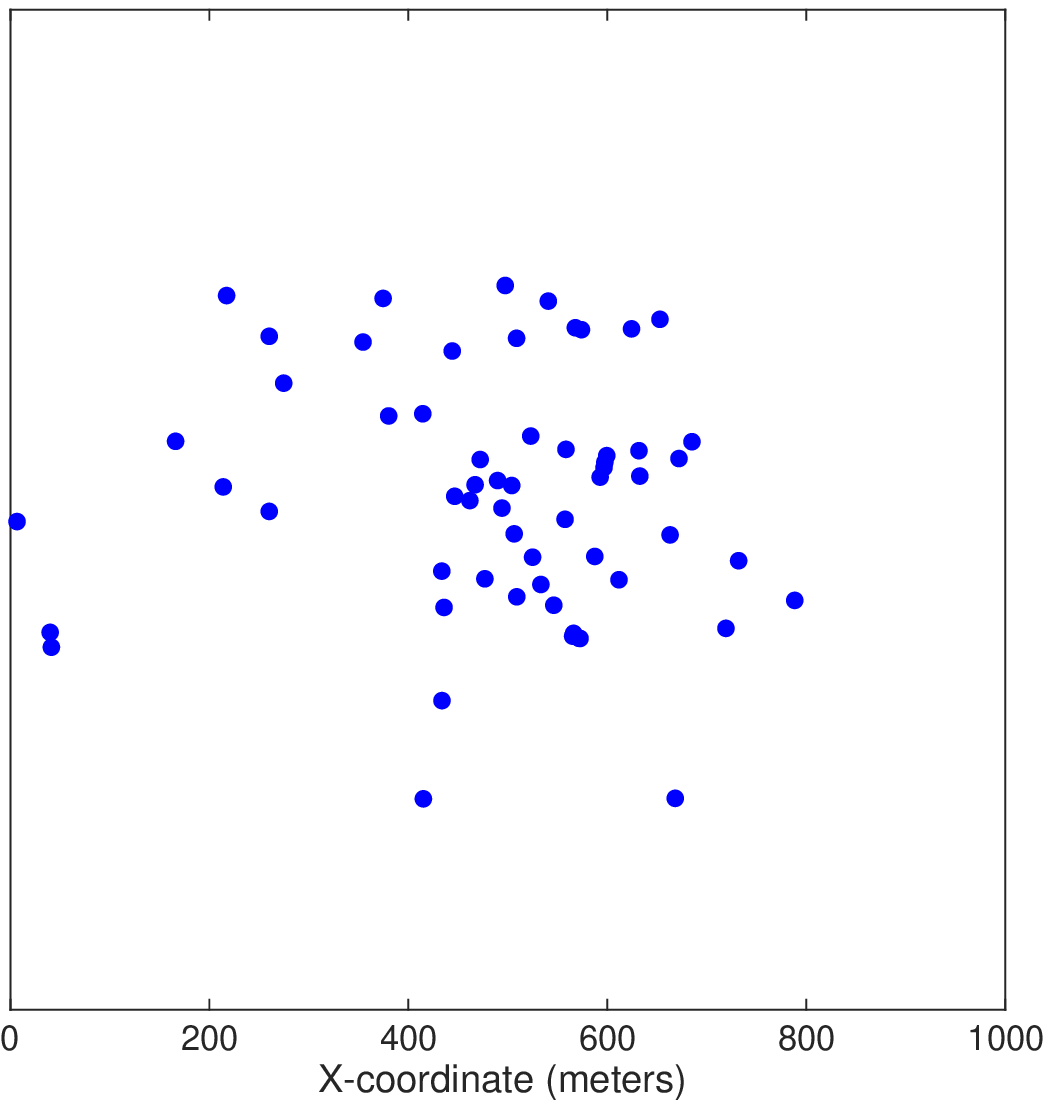}}	
	\subfloat[]{\label{CoV4}\includegraphics[width=1.06in]{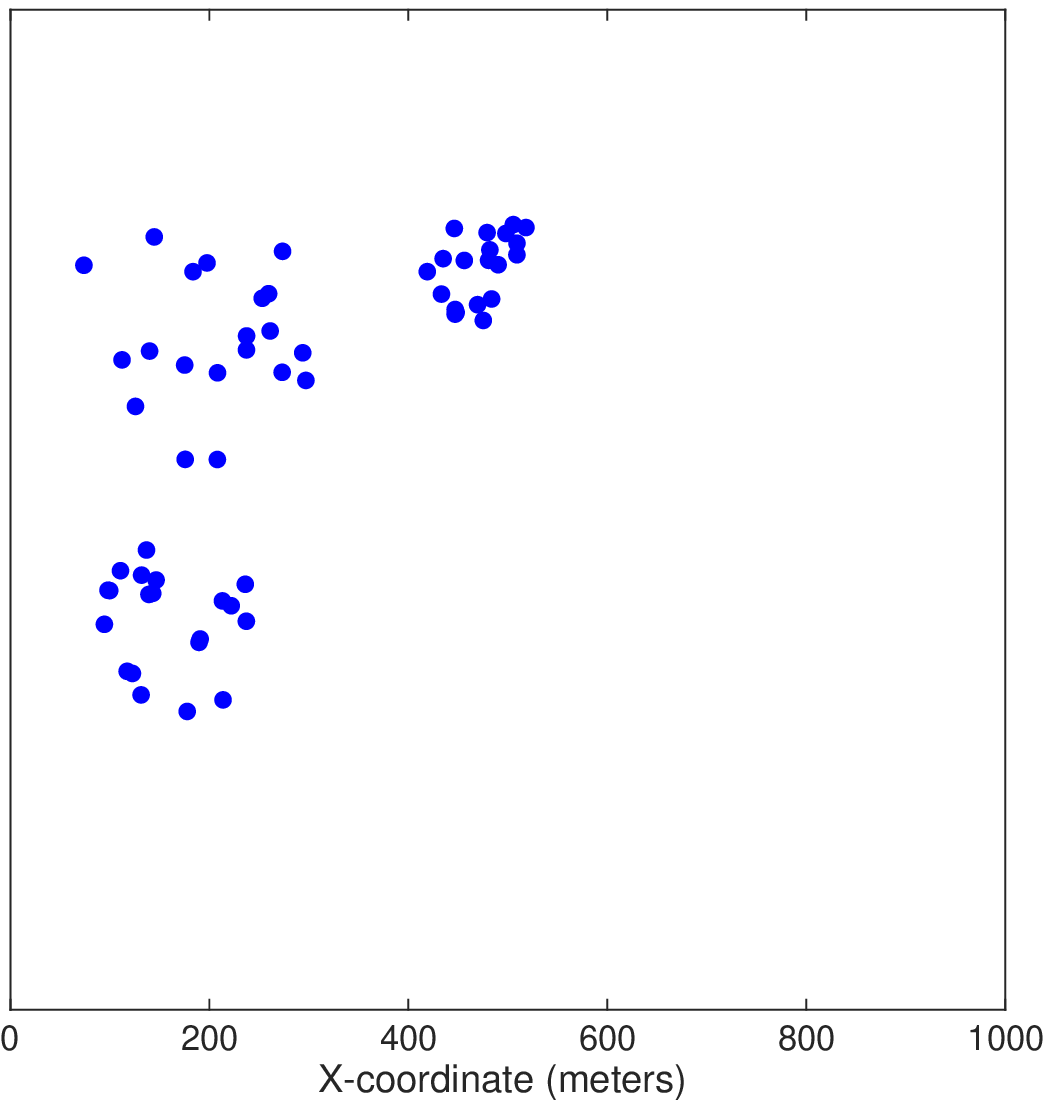}}
	\caption {Sample user distributions with different CoV values, a) CoV=1, b) CoV=2, and c) CoV=4.}
	\label{DifCoV}
\end{figure}

The heterogeneity of user distribution can be measured by the CoV of the Voronoi area of the users \cite{7425184}. CoV is a scalar metric that measures the regularity of the user locations. It is defined as $\frac{1}{0.529} \frac{\sigma_V}{\mu_V}$, where $\sigma_V$ and $\mu_V$ are the standard deviation and the mean of the Voronoi tessellation areas of the users, respectively. CoV=1 corresponds to the Poisson point process, while CoV$>$1 represents clustered distribution of the users. To better visualize different CoV values, Fig.~\ref{DifCoV} shows sample user distributions with their corresponding CoVs. As we can see, higher CoV means that users are more clustered.

\bibliographystyle{IEEEtran}
\bibliography{Xbib}

\begin{thebibliography}{10}
\providecommand{\url}[1]{#1}
\csname url@samestyle\endcsname
\providecommand{\newblock}{\relax}
\providecommand{\bibinfo}[2]{#2}
\providecommand{\BIBentrySTDinterwordspacing}{\spaceskip=0pt\relax}
\providecommand{\BIBentryALTinterwordstretchfactor}{4}
\providecommand{\BIBentryALTinterwordspacing}{\spaceskip=\fontdimen2\font plus
\BIBentryALTinterwordstretchfactor\fontdimen3\font minus
  \fontdimen4\font\relax}
\providecommand{\BIBforeignlanguage}[2]{{%
\expandafter\ifx\csname l@#1\endcsname\relax
\typeout{** WARNING: IEEEtran.bst: No hyphenation pattern has been}%
\typeout{** loaded for the language `#1'. Using the pattern for}%
\typeout{** the default language instead.}%
\else
\language=\csname l@#1\endcsname
\fi
#2}}
\providecommand{\BIBdecl}{\relax}
\BIBdecl

\bibitem{eliVTC}
E.~Kalantari, H.~Yanikomeroglu, and A.~Yongacoglu, ``On the number and {3D}
  placement of drone base stations in wireless cellular networks,'' in
  \emph{2016 IEEE 84th Vehicular Technology Conference (VTC Fall)}, Sep. 2016,
  pp. 1--6.

\bibitem{iremmag}
I.~Bor-Yaliniz and H.~Yanikomeroglu, ``The new frontier in {RAN} heterogeneity:
  Multi-tier drone-cells,'' \emph{IEEE Communications Magazine}, vol.~54,
  no.~11, pp. 48--55, Nov. 2016.

\bibitem{7470933}
Y.~{Zeng}, R.~{Zhang}, and T.~J. {Lim}, ``Wireless communications with unmanned
  aerial vehicles: Opportunities and challenges,'' \emph{IEEE Communications
  Magazine}, vol.~54, no.~5, pp. 36--42, May 2016.

\bibitem{8660516}
M.~{Mozaffari}, W.~{Saad}, M.~{Bennis}, Y.~{Nam}, and M.~{Debbah}, ``A tutorial
  on {UAVs} for wireless networks: Applications, challenges, and open
  problems,'' \emph{IEEE Communications Surveys Tutorials}, vol.~21, no.~3, pp.
  2334--2360, Thirdquarter 2019.

\bibitem{7470932}
S.~Chandrasekharan, K.~Gomez, A.~Al-Hourani, S.~Kandeepan, T.~Rasheed,
  L.~Goratti, L.~Reynaud, D.~Grace, I.~Bucaille, T.~Wirth, and S.~Allsopp,
  ``Designing and implementing future aerial communication networks,''
  \emph{IEEE Communications Magazine}, vol.~54, no.~5, pp. 26--34, May 2016.

\bibitem{7484835}
W.~{Zafar} and B.~{Muhammad Khan}, ``Flying ad-hoc networks: Technological and
  social implications,'' \emph{IEEE Technology and Society Magazine}, vol.~35,
  no.~2, pp. 67--74, Jun. 2016.

\bibitem{8579209}
B.~{Li}, Z.~{Fei}, and Y.~{Zhang}, ``{UAV} communications for {5G} and beyond:
  Recent advances and future trends,'' \emph{IEEE Internet of Things Journal},
  vol.~6, no.~2, pp. 2241--2263, Apr. 2019.

\bibitem{UxNB}
\BIBentryALTinterwordspacing
3GPP, ``3rd generation partnership project; technical specification group
  services and system aspects; unmanned aerial system {(UAS)} support in
  {3GPP},'' {3rd Generation Partnership Project (3GPP)}, Technical
  Specification (TS) 22.125, Dec. 2019, version 17.1.0. [Online]. Available:
  \url{https://portal.3gpp.org/desktopmodules/Specifications/SpecificationDetails.aspx?specificationId=3545}
\BIBentrySTDinterwordspacing

\bibitem{IremTWC}
I.~{Bor-Yaliniz}, A.~{El-Keyi}, and H.~{Yanikomeroglu}, ``Spatial configuration
  of agile wireless networks with drone-{BSs} and user-in-the-loop,''
  \emph{IEEE Transactions on Wireless Communications}, vol.~18, no.~2, pp.
  753--768, Feb. 2019.

\bibitem{AlzenadLetter}
M.~Alzenad, A.~El-Keyi, F.~Lagum, and H.~Yanikomeroglu, ``3-{D} placement of an
  unmanned aerial vehicle base station {(UAV-BS)} for energy-efficient maximal
  coverage,'' \emph{IEEE Wireless Communications Letters}, vol.~6, no.~4, pp.
  434--437, Aug. 2017.

\bibitem{8648498}
Y.~{Sun}, D.~{Xu}, D.~W.~K. {Ng}, L.~{Dai}, and R.~{Schober}, ``Optimal
  {3D}-trajectory design and resource allocation for solar-powered {UAV}
  communication systems,'' \emph{IEEE Transactions on Communications}, vol.~67,
  no.~6, pp. 4281--4298, Jun. 2019.

\bibitem{8377340}
R.~{Ghanavi}, E.~{Kalantari}, M.~{Sabbaghian}, H.~{Yanikomeroglu}, and
  A.~{Yongacoglu}, ``Efficient {3D} aerial base station placement considering
  users mobility by reinforcement learning,'' in \emph{2018 IEEE Wireless
  Communications and Networking Conference (WCNC)}, Apr. 2018, pp. 1--6.

\bibitem{8610014}
J.~{Plachy}, Z.~{Becvar}, P.~{Mach}, R.~{Marik}, and M.~{Vondra}, ``Joint
  positioning of flying base stations and association of users:
  Evolutionary-based approach,'' \emph{IEEE Access}, vol.~7, pp.
  11\,454--11\,463, 2019.

\bibitem{8533634}
M.~{Mozaffari}, A.~{Taleb Zadeh Kasgari}, W.~{Saad}, M.~{Bennis}, and
  M.~{Debbah}, ``Beyond {5G} with {UAVs}: Foundations of a {3D} wireless
  cellular network,'' \emph{IEEE Transactions on Wireless Communications},
  vol.~18, no.~1, pp. 357--372, Jan. 2019.

\bibitem{7841993}
M.~{Mozaffari}, W.~{Saad}, M.~{Bennis}, and M.~{Debbah}, ``Mobile {Internet} of
  things: Can {UAV}s provide an energy-efficient mobile architecture?'' in
  \emph{2016 IEEE Global Communications Conference (GLOBECOM)}, Dec 2016, pp.
  1--6.

\bibitem{azizi}
\BIBentryALTinterwordspacing
A.~Azizi and N.~Mokari, ``Joint resource allocation, {3D} placement, and user
  association in {ABS}-supported {IoT} networks considering adaptive modulation
  technique,'' \emph{Transactions on Emerging Telecommunications Technologies},
  2019. [Online]. Available:
  \url{https://onlinelibrary.wiley.com/doi/abs/10.1002/ett.3632}
\BIBentrySTDinterwordspacing

\bibitem{eliICC}
E.~{Kalantari}, M.~Z. {Shakir}, H.~{Yanikomeroglu}, and A.~{Yongacoglu},
  ``Backhaul-aware robust {3D} drone placement in {5G+} wireless networks,'' in
  \emph{2017 IEEE International Conference on Communications Workshops (ICC
  Workshops)}, May 2017, pp. 109--114.

\bibitem{eliPIMRC}
E.~{Kalantari}, I.~{Bor-Yaliniz}, A.~{Yongacoglu}, and H.~{Yanikomeroglu},
  ``User association and bandwidth allocation for terrestrial and aerial base
  stations with backhaul considerations,'' in \emph{2017 IEEE 28th Annual
  International Symposium on Personal, Indoor, and Mobile Radio Communications
  (PIMRC)}, Oct. 2017, pp. 1--6.

\bibitem{8482444}
T.~M. {Nguyen}, W.~{Ajib}, and C.~{Assi}, ``A novel cooperative {NOMA} for
  designing {UAV}-assisted wireless backhaul networks,'' \emph{IEEE Journal on
  Selected Areas in Communications}, vol.~36, no.~11, pp. 2497--2507, Nov.
  2018.

\bibitem{9014076}
M.~D. {Nguyen}, T.~M. {Ho}, L.~B. {Le}, and A.~{Girard}, ``{UAV} placement and
  bandwidth allocation for {UAV} based wireless networks,'' in \emph{2019 IEEE
  Global Communications Conference (GLOBECOM)}, Dec 2019, pp. 1--6.

\bibitem{8755983}
A.~{Fouda}, A.~S. {Ibrahim}, .~{Güvenç}, and M.~{Ghosh}, ``Interference
  management in {UAV}-assisted integrated access and backhaul cellular
  networks,'' \emph{IEEE Access}, vol.~7, pp. 104\,553--104\,566, 2019.

\bibitem{9042882}
Y.~{Hu}, M.~{Chen}, and W.~{Saad}, ``Joint access and backhaul resource
  management in satellite-drone networks: A competitive market approach,''
  \emph{IEEE Transactions on Wireless Communications}, pp. 1--1, 2020.

\bibitem{7894280}
M.~{Shafi}, A.~F. {Molisch}, P.~J. {Smith}, T.~{Haustein}, P.~{Zhu}, P.~{De
  Silva}, F.~{Tufvesson}, A.~{Benjebbour}, and G.~{Wunder}, ``{5G}: A tutorial
  overview of standards, trials, challenges, deployment, and practice,''
  \emph{IEEE Journal on Selected Areas in Communications}, vol.~35, no.~6, pp.
  1201--1221, Jun. 2017.

\bibitem{wong_schober_ng_wang_2017}
V.~Wong, R.~Schober, D.~Ng, and L.~Wang, \emph{Key Technologies for {5G}
  Wireless Systems}.\hskip 1em plus 0.5em minus 0.4em\relax Cambridge
  University Press, 2017.

\bibitem{8119518}
T.~X. {Vu}, S.~{Chatzinotas}, B.~{Ottersten}, and T.~Q. {Duong}, ``Energy
  minimization for cache-assisted content delivery networks with wireless
  backhaul,'' \emph{IEEE Wireless Communications Letters}, vol.~7, no.~3, pp.
  332--335, Jun. 2018.

\bibitem{8603721}
N.~{Zhao}, F.~R. {Yu}, L.~{Fan}, Y.~{Chen}, J.~{Tang}, A.~{Nallanathan}, and
  V.~C.~M. {Leung}, ``Caching unmanned aerial vehicle-enabled small-cell
  networks: Employing energy-efficient methods that store and retrieve popular
  content,'' \emph{IEEE Vehicular Technology Magazine}, vol.~14, no.~1, pp.
  71--79, Mar. 2019.

\bibitem{Cha2007}
M.~Cha, H.~Kwak, P.~Rodriguez, Y.-Y. Ahn, and S.~Moon, ``I tube, you tube,
  everybody tubes: Analyzing the world's largest user generated content video
  system,'' in \emph{7th ACM SIGCOMM Conference on Internet Measurement (IMC)},
  Oct. 2007, pp. 1--14.

\bibitem{7875131}
M.~{Chen}, M.~{Mozaffari}, W.~{Saad}, C.~{Yin}, M.~{Debbah}, and C.~S. {Hong},
  ``Caching in the sky: Proactive deployment of cache-enabled unmanned aerial
  vehicles for optimized quality-of-experience,'' \emph{IEEE Journal on
  Selected Areas in Communications}, vol.~35, no.~5, pp. 1046--1061, May 2017.

\bibitem{8717714}
M.~{Chen}, W.~{Saad}, and C.~{Yin}, ``Echo-liquid state deep learning for 360
  content transmission and caching in wireless {VR} networks with
  cellular-connected {UAVs},'' \emph{IEEE Transactions on Communications},
  vol.~67, no.~9, pp. 6386--6400, Sep. 2019.

\bibitem{8614433}
------, ``Liquid state machine learning for resource and cache management in
  {LTE-U} unmanned aerial vehicle {(UAV)} networks,'' \emph{IEEE Transactions
  on Wireless Communications}, vol.~18, no.~3, pp. 1504--1517, 2019.

\bibitem{8576651}
B.~{Jiang}, J.~{Yang}, H.~{Xu}, H.~{Song}, and G.~{Zheng}, ``Multimedia data
  throughput maximization in {Internet}-of-things system based on optimization
  of cache-enabled {UAV},'' \emph{IEEE Internet of Things Journal}, vol.~6,
  no.~2, pp. 3525--3532, Apr. 2019.

\bibitem{8387202}
L.~{Wang}, K.~{Wong}, S.~{Jin}, G.~{Zheng}, and R.~W. {Heath}, ``A new look at
  physical layer security, caching, and wireless energy harvesting for
  heterogeneous ultra-dense networks,'' \emph{IEEE Communications Magazine},
  vol.~56, no.~6, pp. 49--55, Jun. 2018.

\bibitem{6193511}
C.~{Fricker}, P.~{Robert}, J.~{Roberts}, and N.~{Sbihi}, ``Impact of traffic
  mix on caching performance in a content-centric network,'' in \emph{2012 IEEE
  INFOCOM Workshops}, Mar. 2012, pp. 310--315.

\bibitem{7386685}
N.~Wang, E.~Hossain, and V.~K. Bhargava, ``Joint downlink cell association and
  bandwidth allocation for wireless backhauling in two-tier {HetNets} with
  large-scale antenna arrays,'' \emph{IEEE Transactions on Wireless
  Communications}, vol.~15, no.~5, pp. 3251--3268, May 2016.

\bibitem{7037248}
A.~Al-Hourani, S.~Kandeepan, and A.~Jamalipour, ``Modeling air-to-ground path
  loss for low altitude platforms in urban environments,'' in \emph{IEEE Global
  Communications Conference (GLOBECOM)}, Dec. 2014, pp. 2898--2904.

\bibitem{7486987}
M.~Mozaffari, W.~Saad, M.~Bennis, and M.~Debbah, ``Efficient deployment of
  multiple unmanned aerial vehicles for optimal wireless coverage,'' \emph{IEEE
  Communications Letters}, vol.~20, no.~8, pp. 1647--1650, Aug. 2016.

\bibitem{Balanis}
C.~A. Balanis, \emph{Antenna Theory: Analysis and Design}.\hskip 1em plus 0.5em
  minus 0.4em\relax Wiley-Interscience, 2005.

\bibitem{3GPP1}
3GPP, ``Selection procedures for the choice of radio transmission technologies
  of the {UMTS},'' {3rd Generation Partnership Project (3GPP)}, Technical
  Report (TR) 30.03U, March. 1998, version 3.2.0.

\bibitem{6834753}
M.~R. {Akdeniz}, Y.~{Liu}, M.~K. {Samimi}, S.~{Sun}, S.~{Rangan}, T.~S.
  {Rappaport}, and E.~{Erkip}, ``Millimeter wave channel modeling and cellular
  capacity evaluation,'' \emph{IEEE Journal on Selected Areas in
  Communications}, vol.~32, no.~6, pp. 1164--1179, June 2014.

\bibitem{7446253}
S.~{Buzzi}, C.~{I}, T.~E. {Klein}, H.~V. {Poor}, C.~{Yang}, and A.~{Zappone},
  ``A survey of energy-efficient techniques for {5G} networks and challenges
  ahead,'' \emph{IEEE Journal on Selected Areas in Communications}, vol.~34,
  no.~4, pp. 697--709, Apr. 2016.

\bibitem{5447068}
Z.~{Luo}, W.~{Ma}, A.~M. {So}, Y.~{Ye}, and S.~{Zhang}, ``Semidefinite
  relaxation of quadratic optimization problems,'' \emph{IEEE Signal Processing
  Magazine}, vol.~27, no.~3, pp. 20--34, May 2010.

\bibitem{boydd}
\BIBentryALTinterwordspacing
J.~Park and S.~Boyd, ``General heuristics for nonconvex quadratically
  constrained quadratic programming,'' Mar. 2017. [Online]. Available:
  \url{https://arxiv.org/abs/1703.07870}
\BIBentrySTDinterwordspacing

\bibitem{Boyd}
S.~Boyd and L.~Vandenberghe, \emph{Convex Optimization}.\hskip 1em plus 0.5em
  minus 0.4em\relax New York, NY, USA: Cambridge University Press, 2004.

\bibitem{cvx}
M.~Grant and S.~Boyd, ``{CVX}: Matlab software for disciplined convex
  programming, version 2.1,'' \url{http://cvxr.com/cvx}, Mar. 2017.

\bibitem{7206589}
L.~{Chen}, F.~R. {Yu}, H.~{Ji}, G.~{Liu}, and V.~C.~M. {Leung}, ``Distributed
  virtual resource allocation in small-cell networks with full-duplex
  self-backhauls and virtualization,'' \emph{IEEE Transactions on Vehicular
  Technology}, vol.~65, no.~7, pp. 5410--5423, Jul. 2016.

\bibitem{mosek}
M.~ApS, \emph{The {MOSEK} optimization toolbox for {MATLAB} Version 9.1.13},
  Feb. 2020.

\bibitem{8038863}
H.~U. {Sokun}, R.~H. {Gohary}, and H.~{Yanikomeroglu}, ``A novel approach for
  {QoS}-aware joint user association, resource block and discrete power
  allocation in {HetNets},'' \emph{IEEE Transactions on Wireless
  Communications}, vol.~16, no.~11, pp. 7603--7618, Nov. 2017.

\bibitem{martin-haenggi}
M.~Haenggi, \emph{Stochastic Geometry for Wireless Networks}.\hskip 1em plus
  0.5em minus 0.4em\relax New York, NY, USA: Cambridge University Press, 2012.

\bibitem{7425184}
F.~Lagum, S.~S. Szyszkowicz, and H.~Yanikomeroglu, ``{CoV}-based metrics for
  quantifying the regularity of hard-core point processes for modeling base
  station locations,'' \emph{IEEE Wireless Communications Letters}, vol.~5,
  no.~3, pp. 276--279, Jun. 2016.

\end{thebibliography}

\begin{IEEEbiography}[{\includegraphics[width=1in,height=1.25in,clip,keepaspectratio]{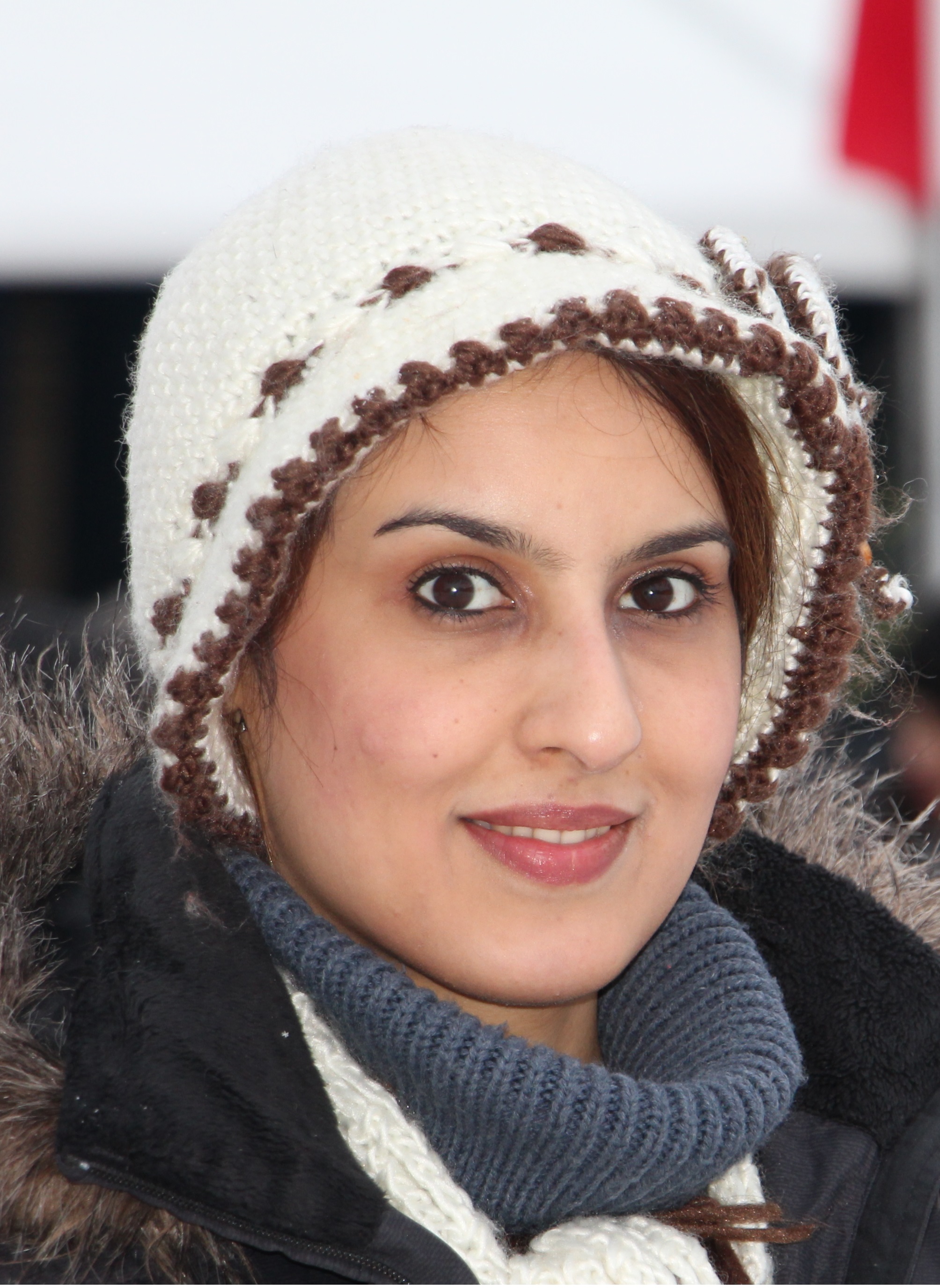}}]{Elham Kalantari}
	received the Ph.D. degree in electrical and computer engineering from university of Ottawa in 2020. She was the recipient of Ontario Graduate Scholarship in 2019. Her research interests are in the areas of optimization, machine learning, resource management, O-RAN, and aerial networks. 
\end{IEEEbiography}

\begin{IEEEbiography}[{\includegraphics[width=1in,height=1.25in,clip,keepaspectratio]{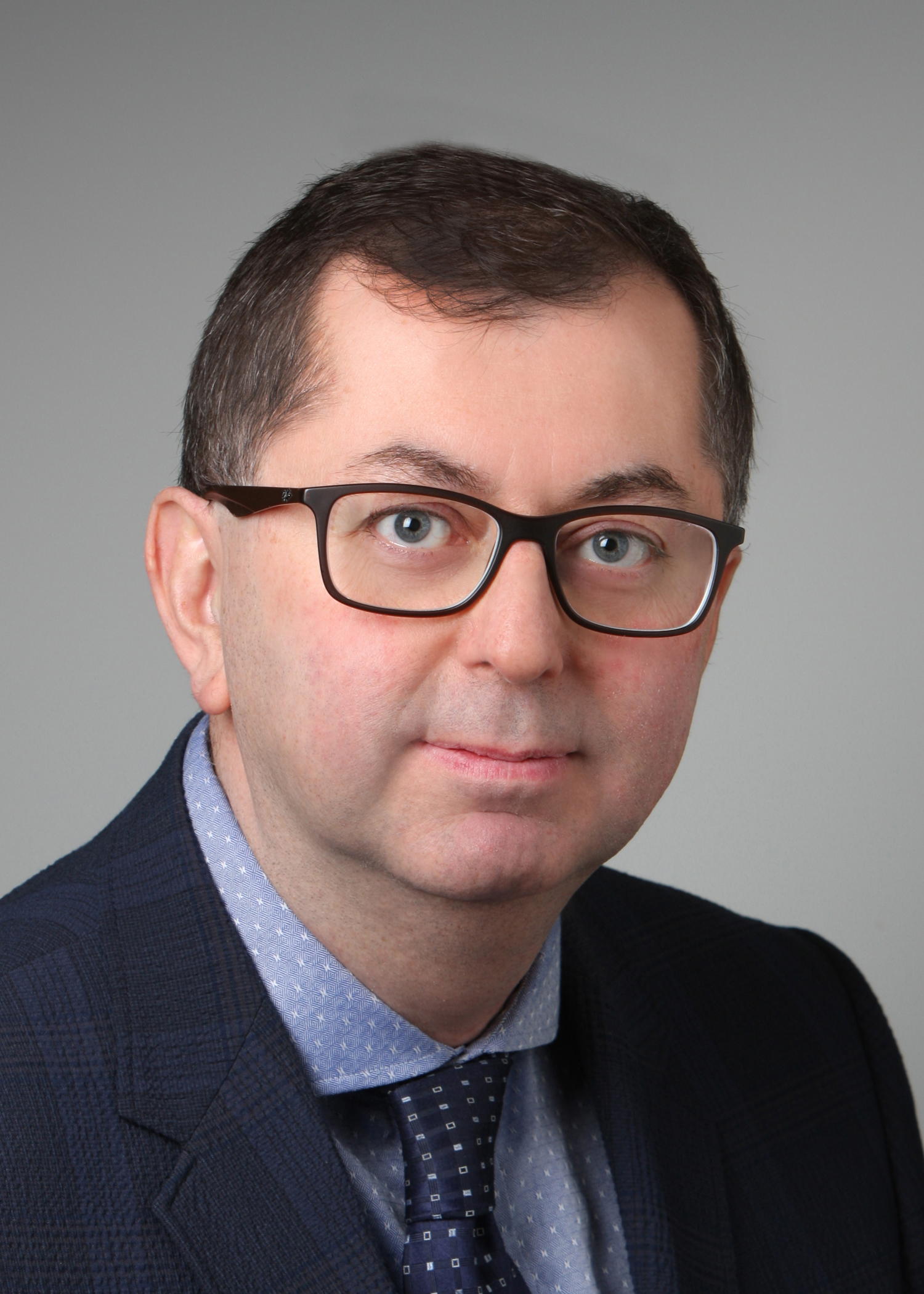}}]{Halim Yanikomeroglu}
	[F] is a professor in the Department of Systems and Computer Engineering at Carleton University, Ottawa, Canada. His research interests cover many aspects of 5G/6G wireless networks. His collaborative research with industry has resulted in 37 granted patents. He is a Fellow of IEEE, Engineering Institute of Canada (EIC), and Canadian Academy of Engineering (CAE); he is a Distinguished Speaker for IEEE Communications Society and IEEE Vehicular Technology Society. He is currently serving as the Chair of the IEEE WCNC (Wireless Communications and Networking Conference) Steering Committee. He was the Technical Program Chair/Co-Chair of WCNC 2004 (Atlanta), WCNC 2008 (Las Vegas), and WCNC 2014 (Istanbul). He was the General Chair of IEEE VTC 2010-Fall (Ottawa) and VTC 2017-Fall (Toronto). He also served as the Chair of the IEEE’s Technical Committee on Personal Communications. Dr. Yanikomeroglu received several awards for his research, teaching, and service, including the IEEE Communications Society Wireless Communications Technical Committee Recognition Award in 2018 and IEEE Vehicular Technology Society Stuart Meyer Memorial Award in 2020.
\end{IEEEbiography}

\begin{IEEEbiography}[{\includegraphics[width=1in,height=1.25in,clip,keepaspectratio]{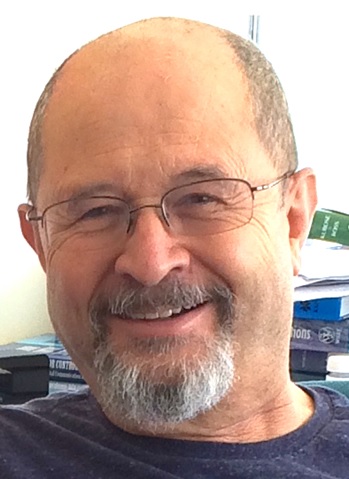}}]{Abbas Yongacoglu}
	received the B.Sc. degree from Bogaziçi University, Turkey, in 1973,  Master's. degree from the University of Toronto in 1975, and the Ph.D. degree from the University of Ottawa in 1987, all in electrical engineering. 
	He worked as a researcher and a system engineer at TUBITAK Marmara Research Institute in Turkey, Philips Research Labs in Holland and Miller Communications Systems in Ottawa. In 1987 he joined the University of Ottawa as an assistant professor. He became an associate professor in 1992, and a full professor in 1996. Since 2018, he is an Emeritus Professor at the University of Ottawa.
	His area of research is wireless communications and digital signal processing. He has been active in organizing several workshops and international conferences on communications and signal processing. He is a Life Senior Member of IEEE.
	
\end{IEEEbiography}

\end{document}